\documentclass[a4paper,oneside,11pt,notitlepage]{report}

\usepackage{geometry}
\usepackage[english]{babel} 
\usepackage[T1]{fontenc}
\usepackage[ansinew]{inputenc}
\usepackage{lmodern} 
\usepackage{eurosym}
\usepackage{cancel}
\usepackage{nameref} 

\usepackage{graphicx} 
\usepackage[position=top]{subfig}   
\usepackage{caption}
\usepackage{array}
\newcolumntype{L}[1]{>{\raggedright\let\newline\\\arraybackslash\hspace{0pt}}m{#1}}
\newcolumntype{C}[1]{>{\centering\let\newline\\\arraybackslash\hspace{0pt}}m{#1}}
\newcolumntype{R}[1]{>{\raggedleft\let\newline\\\arraybackslash\hspace{0pt}}m{#1}}

\usepackage{amsmath}
\usepackage{amsthm}
\usepackage{amsfonts}
\usepackage{dsfont}
\usepackage{xcolor}
\usepackage{mathabx}
\usepackage{array}
\usepackage{slashbox}
\usepackage{mathrsfs}
\usepackage{bm}
\usepackage{lscape}
\usepackage{verbatim}
\usepackage{multicol}
\usepackage{multirow}
\usepackage{changepage}
\usepackage{footnote}
\usepackage{makecell}
\usepackage{lscape}
\usepackage[symbol*]{footmisc}
\usepackage{lipsum}
\usepackage[round]{natbib}
\usepackage[bookmarks=false]{hyperref}
\usepackage{breqn}

\newcommand{\bx}{{\bm x}}
\newcommand{\bX}{{\bm X}}

\newcommand{\rhot}{{\rho_{\tau}}}
\newcommand{\E}{{\mathbb{E}}}

\newcommand{\PP}{{\mathbb{P}}}
\newcommand{\RR}{\mathbb{R}}
\newcommand{\ind}{\mathds{1}}
\newcommand{\vt}{\varphi_\tau}
\newcommand{\psit}{\psi_\tau}

\definecolor{dgreen}{rgb}{0,0.5,0}
\definecolor{dblue}{rgb}{0,0,0.9}
\definecolor{dred}{rgb}{0.7,0.0,0.1}
\definecolor{dgold}{rgb}{0.5,0.3,0.0}
\definecolor{dvio}{rgb}{0.6,0.3,0.5}
\definecolor{gray}{rgb}{0.5,0.5,0.5}
\definecolor{myblue}{rgb}{0.137,0.466,0.741}

\DefineFNsymbolsTM{myfnsymbols}{
  \textasteriskcentered *
  \textdagger    \dagger
}
\numberwithin{equation}{section} 
\theoremstyle{plain}
\newtheorem{theorem}{Theorem}[section] 			 

	\makeatletter
	\def\namedlabel#1#2{\begingroup
			#2%
			\def\@currentlabel{#2}%
			\phantomsection\label{#1}\endgroup
	}
	\makeatother 

\hypersetup{colorlinks,citecolor=blue}

\newcounter{DGPcounter}
\newcommand*{\DGPlabel}[1]{\refstepcounter{DGPcounter}\theDGPcounter\label{#1}}
\newcommand*{\DGPref}[1]{\ref{#1}}
\newcommand{\specificthanks}[1]{*}

\title{An Adapted Loss Function for Censored Quantile Regression}
\author{Micka\"{e}l De Backer\thanks{Universit\'{e} catholique de Louvain, Institut de Statistique, Biostatistique et Sciences Actuarielles. Voie du Roman Pays 20, B-1348 Louvain-la-Neuve, Belgium. Corresponding e-mail: mickael.debacker@uclouvain.be.}\and Anouar El Ghouch\footnotemark[1]\and Ingrid Van Keilegom\textsuperscript{\specificthanks{1},}\thanks{KU Leuven, Research Centre for Operations Research and Business Statistics. Naamsestraat 69, 3000 Leuven, Belgium.}}

\begin{document}


\maketitle
\begin{abstract}

In this paper, we study a novel approach for the estimation of quantiles when facing potential right censoring of the responses. Contrary to the existing literature on the subject, the adopted strategy of this paper is to tackle censoring at the very level of the loss function usually employed for the computation of quantiles, the so-called ``check'' function. For interpretation purposes, a simple comparison with the latter reveals how censoring is accounted for in the newly proposed loss function. Subsequently, when considering the inclusion of covariates for conditional quantile estimation, by defining a new general loss function, the proposed methodology opens the gate to numerous parametric, semiparametric and nonparametric modelling techniques. In order to illustrate this statement, we consider the well-studied linear regression under the usual assumption of conditional independence between the true response and the censoring variable. For practical minimization of the studied loss function, we also provide a simple algorithmic procedure shown to yield satisfactory results for the proposed estimator with respect to the existing literature in an extensive simulation study. From a more theoretical prospect, consistency of the estimator for linear regression is obtained using very recent results on non-smooth semiparametric estimation equations with an infinite-dimensional nuisance parameter, while numerical examples illustrate the adequateness of a simple bootstrap procedure for inferential purposes. Lastly, an application to a real dataset is used to further illustrate the validity and finite sample performance of the proposed estimator. 

\end{abstract}

\newpage
\section{Introduction}
Since the pioneering work of \citet{KB78}, quantile regression has become a preeminent substitute to the classical least-squares regression in both theoretical and applied statistics. While mean regression models solely grasp the central behavior of the data, quantile regression allows the analyst to investigate the complete distributional information of the dependence of the response variable on a set of one or more covariates at hand. In that sense, quantile regression represents a fundamental tool in applications where extremes are important, such as clinical trials or environmental studies where upper and lower quantiles levels are of essential concern. An interesting illustration of this may be found for instance in \citet{EKJ08}, where the average intensity of hurricanes is observed to be steady over the years while upper quantiles are shown to be increasing. This information should then crucially be accounted for in any proper risk or environmental model. Further advantages of quantile regression include, amongst others, flexibility to the error distribution and robustness to outlying observations. A book length treatment of the methodology may be found in \citet{K05}.

Existing literature on the estimation of a quantile regression function includes numerous methodologies for fully observed response observations. In practice however, many interesting applications are affected by possible right censoring of the responses due, for instance, to the withdrawal of patients in biomedical studies, or the end of the follow-up period of a clinical trial. As discussed in \citet{KB01}, \citet{KG01} and \citet{P03}, when confronted to such data, quantile regression provides a valuable complement to the commonly used Cox proportional hazards model or the accelerated failure time model, with ease of interpretation, flexibility to possible heterogeneity in the data and robustness counting as primary benefits.  

The introduction of censored quantile regression goes back to \citeauthor{P84} (\citeyear{P84,P86}) in the econometric literature for linear models and the particular case of fixed censoring, where it is assumed that the censoring variable is always observable. For random censoring, as it is mostly the case in survival analysis, the main rationale of the current literature has been so far to take censoring into account through the formulation of synthetic data points or weighting schemes. One such weighting scheme was introduced by \citet{P03}, where the robustness of quantile regression is exploited through the concept of redistribution-of-mass first developed by \citet{E67}. The underlying idea is to redistribute the mass of censored data lying under the quantile of interest to artificial outlying observations to the right, as the contribution of each point to the estimation of the quantile regression only depends on the sign of the residual. However, given that the weights to be redistributed are to be determined by the conditional distribution of the survival time given the covariates, \citeauthor{P03}'s estimation scheme was developed under a restrictive global linearity assumption to simplify the procedure, that is, assuming that the regression function is linear for all quantiles levels in $(0,\tau),$ where $\tau$ is the quantile level of interest. Relaxing this assumption by only assuming linearity at the level of interest, and exploiting the same idea of redistribution-of-mass, \citet{WW09} proposed a locally weighted estimation scheme based on Beran's local Kaplan-Meier estimator for the conditional distribution of the variable of interest instead of \citeauthor{P03}'s global assumption. Nonetheless, the handling of kernel smoothing causes the estimator of \citeauthor{WW09} to be unsuited for the handling of multiple and categorical covariates. This motivated \citet{WWR14} to exploit, in the same methodology, survival trees instead of Beran's estimator for the required conditional distribution estimation. Overall, the redistribution-of-mass methodology proved to be a valuable approach for parametric censored quantile regression and lead to further research such as variable selection (\citet{WZL13}), multiple quantile estimation (\citet{TW15}) and cure rate quantile regression (\citet{WY16}). However, it should be noted that this general weighting strategy relies on an appropriate estimation of the conditional distribution of the variable of interest. As a consequence, by nature, this strategy only seems appropriate for parametric models as any further extensions of the technique to semiparametric or nonparametric regressions would seem questionable given one would have to adopt a preliminary local estimation of the conditional distribution for, in fact, a local estimator of the conditional quantile.     

An alternative popular weighting scheme in the literature is the so-called inverse-censoring-probability, which was adapted to the linear quantile regression through different versions by \citet{YJW95}, \citet{BT02} and \citet{LT13} to cite a few. This technique is, in spirit, similar to the notorious mean regression estimator of \citet{KSVR81}. The simplicity of use and its ease of interpretation engendered further research such as variable selection (\citet{SLZ10}) as well as extensions outside the linear area, such as nonparametric regression (e.g. \citet{EGVK09}), single-index regression (\citet{BEGVK14}) or copula-based regression (\citet{DBEGVK16}) among others. This advantageous simplicity of adaptation of the technique to a nonlinear regression literature is to be contrasted with the redistribution-of-mass approach. However, major shortcomings of this weighting scheme are the need to evaluate a (possibly local) Kaplan-Meier estimator of the censoring distribution in the right tail, for which \citet{Z06} proposed a simple bypass taking the robustness of quantile regression into account, and, more constraining, an efficiency loss in the estimation as roughly only uncensored observations are preserved in the procedure. 

Besides these fruitful weighting schemes, an interesting idea was proposed by \citet{L97} where, rather than studying weighting strategies for the observed data, the idea was to focus on the quantile level and convert the problem of finding the $\tau$-th conditional quantile of the unobserved variable of interest into finding a superior quantile of the actual observed responses. However, no asymptotics were given and the involved iterative procedure lacks theoretical background.

In this paper, we aim at providing a new insight into censored quantile regression by tackling the problem of censoring through an alternative strategy to the above-discussed literature. In fact, instead of handling weighting schemes or substitute quantile levels, we propose to account for censoring at the very level of the loss function used in quantile regression, the so-called ``check'' function. Intuitively, when confronted to right censored datasets, the proposed loss function will penalize more severely underestimation of the value of the regression than the usual check function. The methodology then allows to plainly exploit the information of all the observations at hand, hence avoiding any estimation efficiency loss. However, as a possible inconvenience, the studied loss function is observed to be no longer convex, which then requires the proposal of an efficient algorithmic procedure to appropriately minimize the resulting objective function. In this paper, a simple adjustment of the Majorize-Minimize (MM) algorithm as proposed by \citet{HL00} is therefore examined. Apart from this, it is worth stressing out that by concentrating on a censored version of the check function, the proposed strategy has the potential to engender multiple extensions to the linear context considered here. The latter could for instance concern parametric models and variable selection in penalized regression, but one could also easily accommodate the strategy to nonparametric and semiparametric models. Therefore, the overall objective of this paper is to illustrate this novel general estimation strategy in a simple and well-studied linear context in order to analyze its behavior in comparison to the existing literature, and hence guide any potential further research on the advantages and pitfalls of this strategy. 

The rest of this paper is organised as follows. Motivating the origin of the newly proposed loss function first for sample quantiles, and then illustrating its accommodation for linear regression is the topic of Section \ref{section:methodology}. Investigating further the linear context, consistency of the proposed estimator is obtained in Section \ref{section:prop}. Section \ref{section:minalgo} provides a detailed adaptation of the MM algorithm to practically implement the procedure, and the finite sample performance of the latter is illustrated by means of
Monte Carlo simulations in Section \ref{section:simulation}. Section \ref{section:application} highlights a brief application to real data. Lastly, the proof of consistency stated in Section \ref{section:prop} is deferred to the Appendix.


\section{The Proposed Methodology}\label{section:methodology}

	
\subsection{Quantiles without covariates}\label{sect:sample}
To motivate and illustrate the proposed methodology, we introduce in this section a simple one-dimensional example and consider the problem of finding the $\tau-$th quantile $m_\tau$ of a survival time variable $T$, or some transformation of the latter. For any $\tau \in (0,1)$, the $\tau-$th quantile is defined as $m_{\tau} = \inf \{t: F_{T}(t) \geq \tau \}$, where $F_T$ is the continuous cumulative distribution function (c.d.f.) of $T$. From the seminal work of \citet{KB78}, it is well-known that this problem is equivalent to solving 
\begin{eqnarray}\label{eq:SQ}
m_{\tau} = \arg\min_a  \E\Big[\rhot(a;T)\Big],
\end{eqnarray}
where $\rhot(a;T) = (T-a)(\tau - \ind(T\leq a))$ is the so-called ``check'' function, and $\ind(\cdot)$ is the indicator function. 

Let us suppose now that the variable of interest $T$ is subject to censoring, that is, instead of observing $T$, one only observes $(Y,\Delta)$, where $Y = \min(T,C)$, $\Delta = \ind(T \leq C)$ and $C$ denotes the censoring variable  assumed for now to be independent of $T$, with c.d.f. $G_C$. In this context, finding $m_\tau$ is equivalent to finding the root in $a$ of the equation 
\begin{align}\label{eq:Esp_nul}
 \E \Big[\ind(Y>a) - \widebar{G}_C(a)(1-\tau)\Big]=0, 
\end{align} 
with $\widebar{G}_C(a):= 1-G_C(a) =\PP(C>a)$. For this equivalence to hold, it is required that $T$ is independent of $C$, and that $\widebar{G}_C(m_\tau)>0$. Note that this latter condition, routinely made in survival analysis, also amounts to establishing a natural upper bound for the quantile of interest that can be studied in the presence of censored data. 

Now, starting from \eqref{eq:Esp_nul}, mimicking the reasoning behind the check function $\rhot$ leads us to define the following function in $a$, for a given $y \in \mathbb{R}$ and $G_C$:
$$\vt(a;y,G_C) = \rhot(a;y) - (1-\tau)\int_0^a G_C(s) \,\mathrm ds,$$ 
which is equal to $\int_a^y \left(\ind(y>s) - \widebar{G}_C(s)(1-\tau) \right)\,\mathrm ds$, up to a constant with respect to $a$. Our claim is that the function $\vt$ actually is an extended version of the check function $\rhot$, such that the effect of censoring is handled at the very level of the loss function through a correcting term $(1-\tau)\int_0^a G_C(s) \mathrm ds$. To support this statement, first note that when all observations are complete, the above-defined function trivially boils down to the check function. Next, observe that, for all values of $a\neq y$, $\partial\vt(a;y,G_C)/\partial a = \widebar{G}_C(a)(1-\tau) - \ind(y>a).$ Since $0\leq \widebar{G}_C(a)\leq 1, \, \forall a$, this suggests that for a value $y<\tau_{G_C} = \inf\{t : G_C(t) = 1\}$, $a \mapsto \vt(a;y,G_C)$ is strictly decreasing on $(-\infty,y)$, strictly increasing on $[y,\tau_{G_C})$, and hence with global minimum value at $a=y$, exactly as the check function. Now, searching for the minimum value in $a$ of $\E\big[\vt(a;Y,G_C)\big]$ reveals to be corresponding to the formulation of \eqref{eq:Esp_nul}, as 
$$\frac{\partial}{\partial a}\E\Big[\vt(a;Y,G_C)\Big] = \widebar{G}_C(a)(1-\tau) - \PP(Y>a).$$ 
From these observations, one may conclude that $\vt$ is an adapted version of the check function $\rhot$ accounting for potential incompleteness of observations. Therefore, in the presence of censoring, we define the logical counterpart of \eqref{eq:SQ}: 
\begin{eqnarray}\label{eq:CSQ}
m_{\tau} = \arg\min_a  \E\Big[\vt(a;Y,G_C)\Big].
\end{eqnarray}
Accordingly, based on an i.i.d. sample $(Y_i,\Delta_i), i=1,\ldots,n,$ of $(Y,\Delta)$, and given an estimator $\widehat{G}_C$ of $G_C$, we define a natural estimator of $m_\tau$ as the empirical version of \eqref{eq:CSQ}:
\begin{eqnarray}\label{eq:CSQ_esti}
\widehat{m}_{\tau} = \arg\min_a  \sum_{i=1}^n \vt(a;Y_i,\widehat{G}_C).
\end{eqnarray}
Note that $\widehat{m}_{\tau}$ exploits all observations in the estimation process, regardless of their censoring status. From the above-described reasoning, it also naturally appears that this general loss function may then be extended outside basic sample quantiles to numerous regression models, as will be illustrated next for the linear model.

\begin{figure}[t!]
		 \captionsetup[subfigure]{labelformat=empty}
  	 \centering
     \subfloat[][]{\includegraphics[scale=.3]{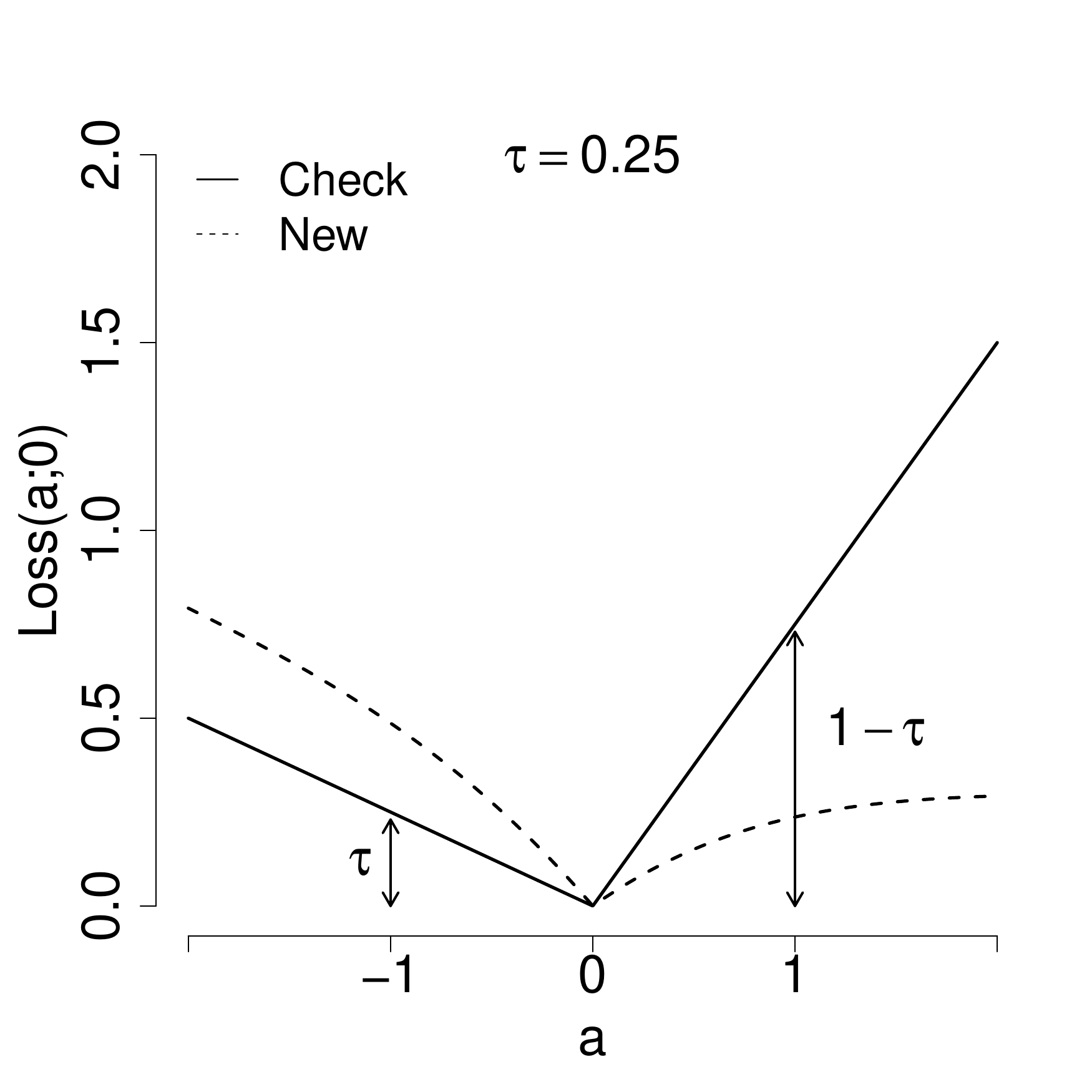}}
     \hspace{0.5cm}
     \subfloat[][]{\includegraphics[scale=.3]{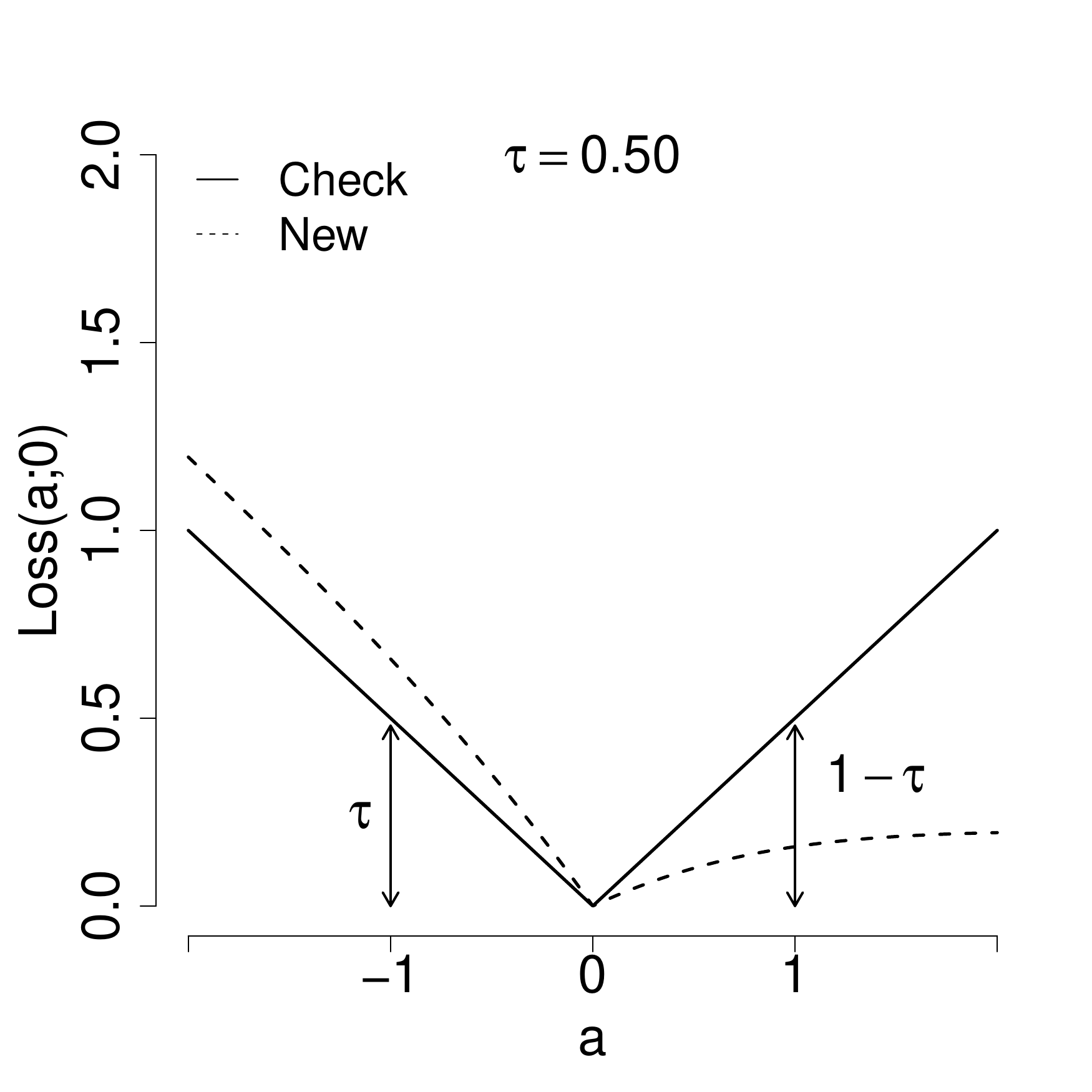}}
		\caption{\small Comparison of the check function $a \mapsto\rhot(a;0)$ with the censored loss function $a \mapsto\vt(a;0,G_C)$ for $\tau=\{0.25,0.5\}$ and $C\sim\mathcal{N}(0,1)$. 
		}
     \label{fig:New_Check}
\end{figure}

To gain further insight into the newly proposed loss function and understand how censoring is accounted for through the regulating term $(1-\tau)\int_0^a G_C(s)\mathrm ds$, we consider in Figure \ref{fig:New_Check} a graphical illustration for the situation $y=0$ with two quantile levels $\tau=\{0.25,0.5\}$ and a standard normal distribution for the censoring variable. Note that in this situation, the correcting term of $\vt$ will always be positive for $a>0$, while negative for $a<0$. Intuitively, this simply suggests that $\vt$ penalizes more heavily underestimation of the value of the quantile than the usual check function to correct for the fact that the response may not be fully observed. Similarly, as can be graphically observed, a smaller loss is assigned for overestimation in order to stimulate the latter when using incomplete data. Note also that the penalization will logically be impacted by the distribution $G_C$, as the amount of alteration of the loss function will be all the more important given that $G_C(m_\tau)=\PP(C\leq m_\tau)$ is large. This is in line with one's intuition that censoring should especially be corrected for when it is susceptible of altering the estimation of a quantile, that is, when censoring is likely to happen to data points lying below $m_\tau$.

Finally, an important feature that should be noted, appearing clearly in Figure \ref{fig:New_Check}, is that $a \mapsto \vt(a;y,G_C)$ is in fact no longer convex. Therefore, special care is called for the numerical optimization required in \eqref{eq:CSQ_esti} as the objective function may allow for local minima, hereby possibly harming the search for the global one. In light of this, we provide in Section \ref{section:minalgo} an adapted version of the Majorize-Minimize (MM) algorithm as proposed by \citet{HL00} that proves to be satisfactory regarding the simulation results of Section \ref{section:simulation}, even in more complex multivariate regression situations.

\subsection{Quantiles with covariates: linear regression}\label{sect:linear regression}
To apply the newly proposed loss function in a regression context, we consider the well-studied linear regression model. Specifically, we consider the inclusion of a covariate vector $\bX$ of dimension $d+1, d\geq 1$, whose first component corresponding to the intercept is taken to be 1. Then, based on an i.i.d. sample $(T_i,\bX_i), i=1,\ldots,n,$ from $(T,\bX)$, it is assumed that
\begin{eqnarray}\label{eq:LQR}
m_{\tau}(\bX_i) = \beta^{\mathsf{T}}_\tau \bX_i,
\end{eqnarray}
where $m_{\tau}(\bx)= \inf \{t: F_{T|\bX}(t|\bx) \geq \tau \}$ is the $\tau-$th conditional quantile of $T$ given $\bX=\bx$, $F_{T|\bX}$ denotes the conditional c.d.f. of $T$ given $\bX$, and $\beta_\tau$ is the $(d+1)-$dimensional unknown quantile coefficient vector. Similarly to sample quantiles, it is well known that the estimation of $\beta_\tau$ in the uncensored case is obtained by minimizing the objective function  
\begin{eqnarray}\label{eq:Unc_esti}
Q_{n}(\beta) = \sum_{i=1}^n \rhot(\beta^{\mathsf{T}} \bX_i;T_i).
\end{eqnarray}

Under random censoring, the statistical problem now consists of estimating the true value of $\beta_\tau$ based on i.i.d triplets $(Y_i,\Delta_i,\bX_i)$, $i=1,\ldots,n,$ from $(Y,\Delta,\bX)$, where, as in Section \ref{sect:sample}, $Y = \min(T,C)$,  $\Delta = \ind(T \leq C)$ and $C$ is assumed here to be independent of $T$ given $\bX$. In this context, analogous arguments to those previously described lead us to define $\beta_\tau$ as the minimizer of the function $$Q^c(\beta) = \E\Big[\vt\left(\beta^{\mathsf{T}} \bX;Y,G_C(\cdot|\bX)\right)\Big],$$ where the superscript $c$ explicitly expresses the presence of censoring, and where $G_C(\cdot|\bx)$ denotes the conditional distribution of $C$ given $\bX=\bx$. Subsequently, we propose to estimate $\beta_\tau$ by minimizing the following objective function:
\begin{eqnarray}\label{eq:Cens_esti}
Q_{n}^c(\beta) = \sum_{i=1}^n \vt\left(\beta^{\mathsf{T}} \bX_i;Y_i,\widehat{G}_C(\cdot|\bX_i)\right) =  \sum_{i=1}^n \left\{\rhot(\beta^{\mathsf{T}} \bX_i;Y_i) - (1-\tau)\int_{0}^{\beta^{\mathsf{T}} \bX_i} \widehat{G}_C(s|\bX_i) \,\mathrm ds\right\}, 
\end{eqnarray}
where $\widehat{G}_C(\cdot|\bx)$ is a consistent estimator of $G_C(\cdot|\bx)$.

Similarly to sample quantiles, this formulation allows one to plainly extract the information of every observation at hand, even if confronted to incompleteness of the latter. In comparison, the initial inverse-censoring-probability strategy of \citet{KSVR81} has to trade-off between appropriate estimation of $G_C$ and efficiency loss through the use of $\Delta$ in the adopted weights. Therefore, one could expect that the proposed methodology will be more efficient than competitors constructed on the basic inverse-censoring-probability strategy, especially when the censoring proportion is large. In the same spirit, one might expect that the above-described estimator provides in this linear context an interesting alternative to the redistribution-of-mass approach when confronted to high censoring percentages, as the latter will inevitably suffer from low sample size for the required estimation of $F_{T|\bX}(\cdot|\bx)$ in this situation. However, a possible drawback of the formulation \eqref{eq:Cens_esti} compared to its competitors is that the estimation of $G_C(\cdot|\bx)$ will affect in our context all observations, hereby possibly harming the global estimation process in case of low sample size for the estimation of the latter distribution. These preliminary remarks will be empirically investigated in Section \ref{section:simulation}.

\section{Large Sample Properties}\label{section:prop}
 
We establish in this section the consistency of the proposed estimator $\widehat{\beta}_\tau$ defined as the minimizer of $Q_{n}^c(\beta)$ in \eqref{eq:Cens_esti}. To that end, we require that the following assumptions hold:   
	\begin{itemize}
	\item[\namedlabel{cond:supp X}{(C1)}] The support supp($\bX$) of $\bX$ is contained in a compact subset of $\RR^{d+1}$, and the variance-covariance matrix of $\bX$ is positive definite.
	\item[\namedlabel{cond:true cond dens}{(C2)}] For $\beta$ in a neighborhood of $\beta_\tau$, $\inf_{\bx \in \text{supp}(\bX)}f_{T|\bX}(\beta^{\mathsf{T}}\bx|\bx)>0$, where $f_{T|\bX}(\cdot|\bx)$ denotes the conditional density function of $T$ given $\bX=\bx$. 
	\item[\namedlabel{cond:True G_C deriv}{(C3)}] Define the (possibly infinite) time $\tau_{F_{Y}(\cdot|\bx)} = \inf\{t : F_{Y|\bX}(t|\bx) = 1\}$, where $F_{Y|\bX}$ designates the conditional c.d.f. of $Y$ given $\bX$. Suppose first that there exists a real number $\upsilon < \tau_{F_{Y}(\cdot|\bx)}$ for all $\bx$ in supp($\bX$). Denote next by $\mathcal{G}$ the class of functions $G(t,\bx):\;]-\infty,\upsilon]\times\text{supp}(\bX) \rightarrow [0,1]$ of bounded variation with respect to $t$ (uniformly in $\bx$) that have first-order partial derivatives with respect to $\bx$ of bounded variation in $t$ (uniformly in $\bx$), and bounded (uniformly in $t$) second-order partial derivatives with respect to $\bx$ which are uniformly in $t$ Lipschitz of order $\eta$ for some $0<\eta<1$. Suppose that $G_C \in \mathcal{G}$.
	\item[\namedlabel{cond:G_C}{(C4)}] For every $\bx \in$ supp$(\bX)$ and for $\beta$ in a neighborhood of $\beta_\tau$, the point $\beta^{\mathsf{T}}\bx \in \RR$ lies below $\upsilon$.
	\item[\namedlabel{cond:Ghat}{(C5)}] The estimator $\widehat{G}_C$ of $G_C$ satisfies $\sup_{\bx \in \text{supp}(\bX)}\sup_{y \leq \upsilon}\left|\widehat{G}_C(y|\bx)-G_C(y|\bx)\right| = o_\PP(1)$, and $\PP(\widehat{G}_C \in \mathcal{G}) \rightarrow 1$ as $n \rightarrow \infty$.
\end{itemize}
Assumptions \ref{cond:supp X} and \ref{cond:true cond dens} are standard in the context of quantile regression estimation for both complete and censored observations in order namely to guarantee the uniqueness of $\beta_\tau$. Assumption \ref{cond:True G_C deriv} defines a general class of functions embedding $G_C$ coming from the work of \citet{L11}. The latter paper develops the theory of bracketing numbers (defined in \citet[p.~83]{VdVW96}) associated to the class $\mathcal{G}$ on which part of the proof relies. Assumption \ref{cond:G_C} is also required for the uniqueness of $\beta_\tau$ and encompasses the remark following equation \eqref{eq:Esp_nul} in Section \ref{sect:sample}, as it defines here in the regression context a natural upper bound for the quantile of interest that can be studied when considering censored responses. Finally, assumption \ref{cond:Ghat} requires the uniform consistency of the censoring distribution estimator and is for instance fulfilled with a conditional, respectively unconditional, Kaplan-Meier estimator as shown in \citet{VKA89}, respectively \citet{G83}, under namely suitable bandwidth conditions for the former.  

\begin{theorem}\label{theorem consist}
For a given quantile level $0<\tau<1$, define $\widehat{\beta}_\tau$ as the minimizer of $Q_{n}^c(\beta)$ in \eqref{eq:Cens_esti}. Assume that the censoring time $C$ is conditionally independent of the survival time $T$ given the covariates $\bX$, and that the triples $(Y_i,\Delta_i,\bX_i), i=1,\ldots,n,$ form an i.i.d. multivariate random sample. Then, under assumptions \ref{cond:supp X}-\ref{cond:Ghat}, 
$$ 
\widehat{\beta}_\tau \rightarrow \beta_\tau
$$ 
in probability, as $n\rightarrow \infty$. 
\end{theorem}
The proof of Theorem \ref{theorem consist}, which is deferred to the Appendix, relies heavily on the work of \citet{DVK15} on non-smooth semiparametric estimating equations with an infinite-dimensional nuisance parameter. The first key requirement to apply this work to our framework is the uniform consistency of the conditional censoring distribution estimator. The second key requirement concerns the class $\mathcal{G}$ for which it has to be shown that the latter is `well-behaved' in terms of size, which is represented by the notion of bracketing number, in order for $\widehat{\beta}_\tau$ to be consistent when constructed on a preliminary estimator $\widehat{G}_C \in \mathcal{G}$. In our proof, we rely on a result of \citet{L11} for the class $\mathcal{G}$ which is constructed upon a dimension reduction technique for the influence of $\bX$ on $G_C$. In particular, \citeauthor{L11} supposes the existence of a function $g: \RR^{d+1} \rightarrow \RR$ for which $G_C(\cdot|\bX)=G_C(\cdot|g(\bX))$. As a consequence, the proof of consistency is here valid as such under the same hypothesis or, more simply, for univariate covariates. 

Showing asymptotic normality of the proposed estimator is a lot less straightforward than consistency, as it requires the establishment of numerous additional technical conditions developed in \citeauthor{DVK15}. Furthermore, as detailed in \citet{K05}, it is known that any conditional quantile regression methodology will depend on the unknown conditional density $f_{T|\bX}$ which will have to be estimated in practice for the covariance matrix of an estimator of $\beta_\tau$. Therefore, for inferential purposes we propose to adopt the simple percentile bootstrap through resampling the triples $(Y_i,\Delta_i,\bX_i), i=1,\ldots,n,$ with replacement. Specifically, 95\% bootstrap confidence intervals for $\beta_\tau$ may easily be constructed using the 2.5th and 97.5th percentiles of the bootstrap coefficients, after having drawn a sufficient amount of bootstrap samples. This technique was repeatedly shown to be satisfactory in the literature (e.g. \citet{P03}, \citet{WW09}, \citet{LT13}) and its validity for the proposed procedure will be illustrated in Section \ref{section:simulation}.

\section{Minimization Algorithm}\label{section:minalgo}
As discussed in Section \ref{section:methodology}, application of the newly proposed methodology for linear regression has to account for the computational aspect of the mathematical minimization of \eqref{eq:Cens_esti}. In particular, in addition to the classical differentiability issue of quantile regression, the nonconvexity of $a \mapsto \vt(a;Y_i,\widehat{G}_C(\cdot|\bX_i))$ requires an efficient optimization routine for the global minimization of the objective function. Given that the newly proposed loss function may be seen as an extension of the check function, we suggest in this section to adapt an existing methodology for uncensored data, for which much work has been provided for the linear context as current procedures include, amongst others, the Interior point algorithm, Simplex algorithms and the MM algorithm. We therefore first briefly review in this section one particular optimization procedure for complete observations, and then discuss how the latter may be adapted to the practical minimization of \eqref{eq:Cens_esti}.

Primarily due to its numerical robustness and ease of adaptation, we propose to investigate in our context the use of the MM, or Majorize-Minimize, algorithm as first introduced by \citet{HL00} for quantile regression. As suggested by its denomination, the rationale of the MM algorithm is to operate in two steps: the objective function to be minimized is first majorized by an appropriate surrogate function, which is then in turn minimized in the second step in order to define the next iterate of the algorithm. By doing so, a difficult optimization problem is to be replaced by a simpler one, with iteration counting as the price to pay for this substitution. More formally, in the quantile regression context with complete observations, letting $\beta^{(m)}$ denote the $m-$th iterate in finding the minimum of $Q_{n}(\beta)$ defined in \eqref{eq:Unc_esti}, \citeauthor{HL00} propose in the first step to majorize $Q_{n}(\beta)$ by a surrogate function $\xi_n(\beta|\beta_{(m)}):\RR^{(d+1)}\times\RR^{(d+1)}\rightarrow\RR$ such that $\xi_n(\beta|\beta_{(m)})\geq Q_{n}(\beta),$ for all $\beta$ and $\xi_n(\beta_{(m)}|\beta_{(m)})=Q_{n}(\beta_{(m)})$. Specifically, exploiting the fact that majorization relations are closed under the formation of sums, \citeauthor{HL00} suggest to majorize $Q_{n}(\beta)$ by the sum for each $i=1,\ldots,n,$ of the unique quadratic curve tangent to the graph of $\rhot(\beta^{\mathsf{T}}_{(m)} \bX_i;T_i)$ at $\beta^{\mathsf{T}}_{(m)} \bX_i$, yielding 
\begin{eqnarray*}
\xi_n(\beta|\beta_{(m)}) = \sum_{i=1}^n \left\{ \frac{(T_i-\beta^{\mathsf{T}} \bX_i)^2}{4(\epsilon+|T_i-\beta^{\mathsf{T}}_{(m)} \bX_i|)} + \left(\tau-\frac{1}{2}\right)(T_i-\beta^{\mathsf{T}} \bX_i) \right\} +c_\tau,
\end{eqnarray*}
where $c_\tau$ is a constant such that $\xi_n(\beta_{(m)}|\beta_{(m)})=Q_{n}(\beta_{(m)})$, and $\epsilon>0$ is a small perturbation to be selected in order to avoid issues with possible zero residuals for iteration $m$. The explicit minimizer of $\xi_n(\beta|\beta_{(m)})$ with respect to $\beta$ then simply becomes the next iterate $\beta_{(m+1)}$ in the second step of the MM algorithm. As an interesting property, called `descent property', it can be shown that the MM algorithm automatically drives the objective function downhill, that is $\xi_n(\beta_{(m+1)}|\beta_{(m)})\leq\xi_n(\beta_{(m)}|\beta_{(m)})$. 

Applying the same arguments with the objective of now finding the minimizer of $Q_{n}^c(\beta)$ defined in \eqref{eq:Cens_esti}, we note that $a \mapsto \vt(a;Y_i,\widehat{G}_C(\cdot|\bX_i))$ is the difference of two convex functions. In the search of a surrogate function, this suggests then to explicitly use the quadratic form of $\xi_n(\beta|\beta_{(m)})$ for the first part of $\vt$, and exploit the convexity of the second part for which we have that for all $i=1,\ldots,n,$ 
$$\int_{0}^{\beta^{\mathsf{T}} \bX_i} \widehat{G}_C(s|\bX_i)\, \mathrm ds \geq  \int_{0}^{\beta^{\mathsf{T}}_{(m)} \bX_i} \widehat{G}_C(s|\bX_i)\, \mathrm ds + (\beta^{\mathsf{T}} \bX_i - \beta^{\mathsf{T}}_{(m)} \bX_i)\widehat{G}_C(\beta^{\mathsf{T}}_{(m)} \bX_i|\bX_i).$$
Recombining this with the expression of $\xi_n(\beta|\beta_{(m)})$ yields the following surrogate function for $Q_{n}^c(\beta)$:  
\begin{multline*}
\xi_n^c(\beta|\beta_{(m)}) =  \sum_{i=1}^n \Bigg\{ \frac{(Y_i-\beta^{\mathsf{T}} \bX_i)^2}{4(\epsilon+|Y_i-\beta^{\mathsf{T}}_{(m)} \bX_i|)} + \left(\tau-\frac{1}{2}\right)(Y_i-\beta^{\mathsf{T}} \bX_i)  \\
- (1-\tau)\int_{0}^{\beta^{\mathsf{T}}_{(m)} \bX_i} \widehat{G}_C(s|\bX_i)\, \mathrm ds - (1-\tau)(\beta^{\mathsf{T}} \bX_i - \beta^{\mathsf{T}}_{(m)} \bX_i)\widehat{G}_C(\beta^{\mathsf{T}}_{(m)} \bX_i|\bX_i) \Bigg \}  + \tilde{c}_\tau,
\end{multline*}
where $\tilde{c}_\tau$ is a constant such that $\xi_n^c(\beta_{(m)}|\beta_{(m)})=Q_{n}^c(\beta_{(m)})$. For the second step of the algorithm, let $\mathcal{X}=[\bX_1,\ldots,\bX_n]$ be the $(d+1)\times n$ matrix of covariates, and $\mathcal{Y}^{\mathsf{T}}=(Y_1,\ldots,Y_n)$ be the vector of observed responses. Minimizing $\xi_n^c(\beta|\beta_{(m)})$ with respect to $\beta$ to obtain $\beta_{(m+1)}$ then yields the following succinct result:
\begin{eqnarray}\label{eq:MM_cens}
- \mathcal{X} \mathcal{A}_{(m)} \mathcal{Y} + \mathcal{X} \mathcal{A}_{(m)} \mathcal{X}^{\mathsf{T}} \beta_{(m+1)} - \mathcal{X} \mathcal{D} = \mathcal{X} \mathcal{E}_{(m)},
\end{eqnarray}
where $\mathcal{A}_{(m)}$ is a $n\times n$ diagonal matrix with $i$-th diagonal entry $1/\big[2(\epsilon-|Y_i-\beta^{\mathsf{T}}_{(m)} \bX_i|)\big]$, $\mathcal{D}$ is a $n\times 1$ vector of $(\tau-1/2)$, and $\mathcal{E}_{(m)}$ is a $n\times 1$ vector with $i$-th entry $(1-\tau)\widehat{G}_C(\beta^{\mathsf{T}}_{(m)} \bX_i|\bX_i)$. Solving \eqref{eq:MM_cens} with respect to $\beta_{(m+1)}$ yields the explicit iteration of the MM algorithm at step $(m+1)$, which is a simple adaptation of the MM algorithm for linear quantile regression with complete observations as there is only a supplementary term $\mathcal{E}_{(m)}$ appearing at every iteration. The proposed algorithm may then be resumed as follows:
\subsubsection*{Algorithm}
\begin{enumerate}
	\item[Step 0.] Given an initial estimate of $\beta_\tau$ denoted by $\beta_{(0)}$, set $m=0$. Select a small tolerance value $\delta$, for instance $\delta=10^{-9}$, and choose $\epsilon$ such that $\epsilon \ln\epsilon\approx-\delta/n$.
	\item[Step 1.] Estimate $G_C(\cdot|\bX)$ and calculate $\mathcal{E}_{(m)}$ as $(1-\tau)\widehat{G}_C(\beta^{\mathsf{T}}_{(m)} \bX_i|\bX_i), i=1,\ldots,n$. Calculate $\mathcal{A}_{(m)}$ and set $$\beta_{(m+1)} = \left(\mathcal{X} \mathcal{A}_{(m)} \mathcal{X}^{\mathsf{T}}\right)^{-1} \mathcal{X}\left(\mathcal{A}_{(m)} \mathcal{Y} + \mathcal{D} + \mathcal{E}_{(m)}\right).$$
	\item[Step 2.] If $||\beta_{(m+1)}-\beta_{(m)}||>\delta$ or $\left|\xi_n^c(\beta_{(m+1)}|\beta_{(m)}) - \xi_n^c(\beta_{(m)}|\beta_{(m)})\right|>\delta$, replace $m$ by $m+1$ and return to step 1.
\end{enumerate} 

Concerning the choice of $\beta_{(0)}$, special care is to be brought to the latter given the nonconvexity of $\vt$, and hence the possibility of capturing only a local minimum, if there is any. In an attempt to mend this obstacle, similarly to \citet{LT13}, we propose to adopt the estimator of \citet{BT02} as $\beta_{(0)}$, given the consistency of the latter. Hence, one could hope that, by already providing a descent estimation of $\beta_\tau$, $\beta_{(0)}$ should be close to the global minimum of $Q_{n}^c(\beta)$, hereby possibly avoiding the ambush of potential local minima. Furthermore, given the need at each iteration $m$ of the algorithm to evaluate $\widehat{G}_C(\beta^{\mathsf{T}}_{(m)} \bX_i|\bX_i), i=1,\ldots,n,$ a descent estimator $\beta_{(0)}$ may also, to some extent, prevent too many evaluations of $\widehat{G}_C(\cdot|\bX)$ in the right tail, for which classical estimators are known to be typically unstable due to sparsity of the data. On the other hand, as noted by \citeauthor{LT13}, given that only uncensored observations are handled for $\beta_{(0)}$, the latter estimator may severely suffer from an efficiency loss in the estimation, hereby motivating the use of an estimator exploiting all observations at hand such as the one proposed in this paper. Nonetheless, as there is no guarantee that the algorithm starting from this $\beta_{(0)}$ will avoid converging to a possible local minimum, it is preferable to restart the latter as suggested by \citeauthor{HL00}, in this case for instance by adding small perturbations to $\beta_{(0)}$ to verify the stability of the solution. 

Finally, note that, analogously to the remarks of Section \ref{section:methodology}, the described algorithmic procedure of this section based on the MM algorithm is again easily adaptable to broader parametric, nonparametric and semiparametric models. For instance, considering a general parametric model, a convenient minimization algorithmic procedure would only require an adaptation of equation \eqref{eq:MM_cens} to define the appropriate iterative process one could readily implement for the desired model.

\section{Simulation Study}\label{section:simulation}
In this section, we assess the finite sample performance of the proposed methodology by means of Monte Carlo simulations. As previously mentioned, given that the adapted loss function may engender further modelling techniques beyond the simple linear framework, we are mainly interested here in comparing the performance of our estimator (NEW) with the two prominent weighting schemes for quantile regression with censored data: the redistribution-of-mass (RM), which will be embodied here by \citeauthor{WW09}'s estimator, and the inverse-censoring-probability (ICP), represented here by \citeauthor{BT02}' estimator. Based on the choice of these competitors, as a preliminary remark that has repeatedly been illustrated in the literature, while ICP may be subject to possible improvement for linear models, RM provides in this context a very competitive estimator, often taken as primary reference. Lastly, to grasp the impact of censoring, we also include an omniscient procedure (\textit{Omni}) using all the observations $T_i, i=1,\ldots,n,$ as if they were available in practice for the estimation of $\beta_\tau$. 

The four procedures are compared in four main data generating processes (DGP) where the survival time is systematically independent of the censoring time given the covariates. The first three DGPs are taken to be univariate settings, while the fourth setting considers an example with four covariates. Given that both RM and NEW will be implemented with local estimators of conditional distributions as will be detailed below, the latter setting will be of interest in order to analyze the performance of both procedures when confronted to multivariate covariates. Note that, in their paper on variable selection using the redistribution-of-mass, \citet{WZL13} also tolerate as many as four covariates for the determination of the involved local weights.

The first DGP is taken from \citet{WW09} and \citet{LT13}, and presents a simple setting which is linear in all quantile levels and where all estimation procedures in the literature perform relatively well for the estimation of the regression coefficients. Hence, a basic requirement for a newly proposed methodology as the one considered here would be to perform appropriately as well in this particular setting. Next, the second and third DGP, partly inspired by \citeauthor{WW09}, are taken to be linear only in the quantile level of interest, hereby ruling out the use of estimators assuming global linearity such as \citeauthor{P03}'s or the martingale-based estimator of \citet{PH08}, despite their relative robustness illustrated for instance in \citeauthor{WW09} and \citeauthor{LT13}. For the second DGP, the censoring variable $C$ is taken to be independent of the covariate, while the third DGP covers a dependent scenario. This will allow to explore the potential impact of a conditional censoring distribution. Note however that, in all four settings, the proposed procedure NEW will repeatedly be implemented considering a conditional censoring distribution even when the simulated settings do not present any influence of the covariate on the latter distribution. 

Concerning the implementation of these procedures, we first use the \texttt{rq} function of the R library \texttt{quantreg} for \textit{Omni}. ICP is likewise implemented with \texttt{rq} by incorporating in the function the weights $\Delta_i/(1-\widehat{G}_C(Y_i)), i=1,\ldots,n,$ where $\widehat{G}_C$ is the Kaplan-Meier estimator of $G_C$. As for RM, the estimator is implemented using \citeauthor{WW09}'s code, available on their websites. For the estimation of local weights, we use the biquadratic kernel $K(x)=(15/16)(1-x^2)^2\ind(|x|\leq 1)$ for the univariate settings as recommended in \citeauthor{WW09}, while an eighth-order kernel $K(x)=(1/13)(1-x^2)(35-385x^2+1001x^4-715x^6)\ind(|x|\leq 1)$ is used for the four-variate example, just as in \citeauthor{WZL13}. For the first three settings, the bandwidth of the procedure is determined by 5-fold cross validation among 15 candidates equally ranging from 0.05 to 0.5, whereas one initial simulated dataset serves for the determination of the bandwidth  for the multivariate DGP among 15 candidates equally ranging from 0.5 to 2 given the computational cost of searching for tuning parameters across multiple dimensions. Lastly, NEW is implemented using the algorithm of Section \ref{section:minalgo} where, for every DGP we adopt Beran's local Kaplan-Meier estimator as $\widehat{G}_C(\cdot|\bX)$. As for RM, the latter is based either on the biquadratic or the eighth-order kernel depending on the dimension of the covariate. Finally, the required bandwidth is likewise computed via 5-fold cross validation at each iteration or on one initial dataset. This suggests that for the last DGP, the performance of both RM and NEW could probably be improved if the bandwidths were adapted to each simulation, although none of the estimators is here favored over the other in the analysis we intend to provide.

\subsubsection*{DGP \DGPlabel{DGP:1}.}
The following model is generated: 
$$T_i = \beta_0 + \beta_1X_{i} + \eta_i,$$ 
where $\beta_0=3$, $\beta_1=5$, $X_1,\ldots,X_n$ are i.i.d. $U[0,1]$ variables and $\eta_1,\ldots,\eta_n$ are i.i.d. $\mathcal{N}(0,1)$ variables. The censoring variables $C_1,\ldots,C_n$ are simulated from $U[0,M]$, with $M$ calibrated to attain the desired censoring proportion (15\% or 40\%) at the median. We note that the choice of $M$ also implies that condition \ref{cond:G_C} is not violated for the true $\beta_\tau$, where $\tau=0.5$ in this case. The latter consideration regarding condition \ref{cond:G_C} repeatedly applies to all the following DGPs and for all the considered levels of $\tau$.
\subsubsection*{DGP \DGPlabel{DGP:2}.} 
The following model is generated: 
$$ T_i = \beta_0 + \beta_1X_{i} + \left(3+(X_{i}-0.5)^2\right)\left(\eta_i - \Phi^{-1}(\tau)\right),$$ 
where $\beta_0=1$, $\beta_1=0.1$, $X_1,\ldots,X_n$ are i.i.d. $\mathcal{N}(0,1)$ variables, $\eta_1,\ldots,\eta_n$ are also i.i.d. $\mathcal{N}(0,1)$ variables and $\Phi^{-1}$ denotes the quantile function of the standard normal distribution. The censoring variables $C_1,\ldots,C_n$ are simulated from $U[m,M],$ where $m$ and $M$ are chosen to attain the desired censoring proportions ($30\%$ and $60\%$) depending on the considered quantile levels. 
\subsubsection*{DGP \DGPlabel{DGP:3}.}
The model to generate $T_i, i=1,\ldots,n,$ is the same as for DGP \DGPref{DGP:2}. However, in this setting the censoring variable is generated from the model $C_i = \gamma_0 + \gamma_1 X_i + \upsilon_i, i=1,\ldots,n,$ where $\gamma_0=1$, $\gamma_1=-0.1$, and $\upsilon_1,\ldots,\upsilon_n$ are i.i.d. variables from $U[m,M]$, where $m$ and $M$ are selected to obtain the same censoring levels as for DGP \DGPref{DGP:2}.
\subsubsection*{DGP \DGPlabel{DGP:4}.}
The following model, inspired by \citet{SLZ10}, is generated: 
$$ T_i = \beta_0 + \beta_1X_{1i} + \beta_2X_{2i} +\beta_3X_{3i} +\beta_4X_{4i} +\eta_i,$$ 
where $\beta_0=1$, $\beta_1=0.5$, $\beta_2=1$, $\beta_3=1.5$, $\beta_4=2$, $X_{ji}$ are i.i.d. standard normal variables for $i=1,\ldots,n,$ and $j=1,\ldots,4$, and $\eta_1,\ldots,\eta_n,$ are i.i.d. variables simulated from $t(5)$. The censoring variables $C_1,\ldots,C_n$ are, here again, simulated from $U[m,M],$ where $m$ and $M$ are chosen to attain the desired censoring proportions ($30\%$ and $60\%$) at the quantile level $\tau=0.5$.  

For all simulation settings, we consider $B=500$ repetitions of each DGP, four average censoring proportions ($p_c \in \{15\%,40\%\}$ for DGP \DGPref{DGP:1} and $p_c \in \{30\%,60\%\}$ for DGP \DGPref{DGP:2}, \DGPref{DGP:3} and \DGPref{DGP:4}), and sample sizes $n \in \{100,200,500\}$. The quantile levels of interest are chosen among $\{0.3,0.5,0.7\}$ depending on the DGP and the censoring proportion of interest. Estimators are compared in terms of bias, root mean squared errors (RMSE), and median absolute error (MAE). We further include an overall criterion taken from a prediction point of view: the mean absolute deviation (MAD), given for an estimator $\widehat{\beta}$ by $n^{-1}\sum_{i=1}^n |\widehat{\beta}^{\mathsf{T}}\bX_i - \beta^{\mathsf{T}}_\tau \bX_i|$.

\begin{table}[t!] 
\small
\centering
\begin{tabular}{ccc||C{1cm}C{1cm}C{1cm}C{1cm}C{1cm}C{1cm}|C{1.1cm}}
\multicolumn{3}{c}{$$}&\multicolumn{2}{c}{Bias}&\multicolumn{2}{c}{RMSE}&\multicolumn{2}{c}{MAE}\\
$n$ & $p_c$ & Method & $\beta_0$ & $\beta_1$ & $\beta_0$ & $\beta_1$ &$\beta_0$ & $\beta_1$&MAD\\
\hline
\hline 
\multirow{8}{*}{100} & \multirow{4}{*}{15\%} & \textit{Omni} & 0.007 & 0.000 & 0.255 & 0.425 & 0.168 & 0.281 & 0.135 \\ [1.1pt]
  &&NEW   & 0.013 & -0.006 & 0.267 & 0.454 & 0.175 & 0.282 & 0.147\\ [1.1pt]
  &&RM    & 0.009 & -0.004 & 0.267 & 0.459 & 0.173 & 0.286 & 0.148\\ [1.1pt]
  &&ICP   & 0.010 & -0.006 & 0.270 & 0.456 & 0.174 & 0.283 & 0.148 \\ [1.1pt]\cline{3-10}			
											  & \multirow{4}{*}{40\%}& \textit{Omni} & 0.007 & 0.000 & 0.255 & 0.425 & 0.168 & 0.281 & 0.135 \\ [1.1pt]
  &&NEW   & 0.009 & 0.013 & 0.298 & 0.558 & 0.183 & 0.343 & 0.177 \\ [1.1pt]
  &&RM    & 0.001 & 0.003 & 0.297 & 0.552 & 0.184 & 0.361 & 0.176 \\ [1.1pt]
  &&ICP   & 0.009 & -0.001 & 0.311 & 0.583 & 0.185 & 0.384 & 0.181 \\ [1.1pt]\cline{3-10}
\hline
\hline
\multirow{8}{*}{200} & \multirow{4}{*}{15\%} & \textit{Omni} & 0.009 & -0.006 & 0.169 & 0.298 & 0.108 & 0.204 & 0.095 \\ [1.1pt]
  &&NEW   & 0.013 & -0.009 & 0.182 & 0.323 & 0.126 & 0.227 & 0.104  \\ [1.1pt]
  &&RM    & 0.011 & -0.011 & 0.184 & 0.325 & 0.127 & 0.230 & 0.105 \\ [1.1pt]
  &&ICP   & 0.013 & -0.014 & 0.182 & 0.322 & 0.129 & 0.226 & 0.104  \\    [1.1pt]\cline{3-10}			
											  & \multirow{4}{*}{40\%}& \textit{Omni} & 0.009 & -0.006 & 0.169 & 0.298 & 0.108 & 0.204 & 0.095 \\ [1.1pt]
  &&NEW   & 0.014 & -0.011 & 0.206 & 0.390 & 0.133 & 0.267 & 0.124  \\ [1.1pt]
  &&RM    & 0.008 & -0.025 & 0.204 & 0.387 & 0.134 & 0.268 & 0.123 \\ [1.1pt]
  &&ICP   & 0.016 & -0.027 & 0.217 & 0.403 & 0.148 & 0.276 & 0.127 \\[1.1pt]\cline{3-10}
\hline
\hline
\end{tabular}
\caption{\small Simulation results for DGP \DGPref{DGP:1} expressed in terms of bias, RMSE, MAE and MAD averaged over $B=500$ repetitions. Average censoring proportions $p_c$ are taken in $\{15\%,40\%\}$ at the median, and sample sizes are taken in $\{100,200\}$. 
}
\label{table:DGP1}
\end{table}	

Table \ref{table:DGP1} reports the results of our simulation study for DGP \DGPref{DGP:1}. As expected, all estimation procedures perform very similarly in this basic setting, although ICP already exhibits some relative difficulties to compete when confronted to higher censoring proportions. Concerning NEW, we note that in this example the procedure is relatively robust to the low sample size for computing the required local Kaplan-Meier estimator, as observed for $n=100$ and $p_c=15\%$. Furthermore, even though the simulated scenario does not account for any influence of the covariate on the censoring distribution, our simulation experience with the present DGP suggests that the estimator is relatively robust to the smoothing parameter to be selected when using Beran's estimator $\widehat{G}_C(\cdot|X)$. 
This robustness of the results towards both the sample size and the smoothing parameter may be considered as an encouraging observation for the newly proposed estimator given that all observations are affected by the conditional censoring distribution estimator through the studied adapted loss function as mentioned in Section \ref{section:methodology}. 


\begin{table}[b!] 
\small
\centering
\begin{tabular}{ccc||C{1cm}C{1cm}C{1cm}C{1cm}C{1cm}C{1cm}|C{1.1cm}}
\multicolumn{3}{c}{$$}&\multicolumn{2}{c}{Bias}&\multicolumn{2}{c}{RMSE}&\multicolumn{2}{c}{MAE}&\\
$p_c$ & $\tau$ & Method & $\beta_0$ & $\beta_1$ & $\beta_0$ & $\beta_1$ &$\beta_0$ & $\beta_1$&MAD\\
\hline
\hline 
\multirow{8}{*}{30\%} & \multirow{4}{*}{0.3} & \textit{Omni} & 0.021 & 0.014 & 0.373 & 0.507 & 0.244 & 0.337 & 0.466 \\ [1.1pt]
  &&NEW & 0.014 & 0.031 & 0.388 & 0.509 & 0.258 & 0.350 & 0.479 \\ [1.1pt]
  &&RM & -0.116 & 0.057 & 0.400 & 0.486 & 0.265 & 0.321 & 0.473 \\ [1.1pt] 
  &&ICP & -0.071 & 0.140 & 0.400 & 0.582 & 0.264 & 0.363 & 0.513 \\ [1.1pt]\cline{3-10}			
												& \multirow{4}{*}{0.7}& \textit{Omni} & 0.014 & -0.013 & 0.371 & 0.502 & 0.243 & 0.332 & 0.456 \\[1.1pt]
  &&NEW & -0.176 & 0.049 & 0.445 & 0.502 & 0.293 & 0.324 & 0.538 \\ [1.1pt]
  &&RM & -0.403 & 0.104 & 0.560 & 0.456 & 0.404 & 0.311 & 0.581 \\ [1.1pt] 
  &&ICP & -0.317 & 0.218 & 0.570 & 0.732 & 0.417 & 0.504 & 0.691 \\  [1.1pt]\cline{3-10}
\hline
\hline
\multirow{8}{*}{60\%} & \multirow{4}{*}{0.3} & \textit{Omni} & 0.021 & 0.014 & 0.373 & 0.507 & 0.244 & 0.337 & 0.466 \\ [1.1pt]
  &&NEW & -0.058 & 0.075 & 0.390 & 0.503 & 0.268 & 0.338 & 0.473 \\ [1.1pt]
  &&RM & -0.517 & 0.102 & 0.627 & 0.414 & 0.524 & 0.290 & 0.613 \\ [1.1pt] 
  &&ICP & -1.097 & 0.442 & 1.208 & 0.792 & 1.083 & 0.502 & 1.204 \\  [1.1pt]\cline{3-10}			
											  & \multirow{4}{*}{0.5}& \textit{Omni} & 0.002 & 0.009 & 0.350 & 0.473 & 0.226 & 0.308 & 0.432 \\[1.1pt] 
  &&NEW & -0.269 & 0.113 & 0.504 & 0.564 & 0.339 & 0.353 & 0.565 \\[1.1pt] 
  &&RM & -0.744 & 0.190 & 0.828 & 0.475 & 0.760 & 0.346 & 0.815 \\ [1.1pt] 
  &&ICP & -1.387 & 0.477 & 1.487 & 0.856 & 1.395 & 0.566 & 1.486 \\    [1.1pt]\cline{3-10}
\hline
\hline
\end{tabular}
\caption{\small Simulation results for DGP \DGPref{DGP:2} with sample size $n=200$, expressed in terms of bias, RMSE, MAE and MAD averaged over $B=500$ repetitions. Average censoring proportions $p_c$ are taken in $\{30\%,60\%\}$, and quantile levels $\tau$ are chosen among $\{0.3,0.5,0.7\}$.
}
\label{table:Ex1n200}
\end{table}

Moving to a more challenging setting with higher censoring proportions, Table \ref{table:Ex1n200} reports the results of our simulation study for DGP \DGPref{DGP:2} with $n=200$ observations at each iteration. Due to identifiability issues as encountered in condition \ref{cond:G_C}, the quantile level $\tau=0.7$ is not considered here when simulating as much as $60\%$ censoring, whereas the quantile level $\tau=0.5$ is left aside for $30\%$ censoring for the sake of brevity. From these results, we observe that both RM and ICP present substantial biases in the parameter estimation for this DGP, particularly for the intercept. In contrast, NEW exhibits difficulties in terms of bias only for the most complicated situations considered here, that is, for high quantile levels with respect to the censoring proportion. These bias results concerning the estimation of $\beta_0$ naturally impact the RMSE and MAE of both RM and ICP. However, while ICP presents considerable biases for both parameters to be estimated, RM quite surprisingly seems to compensate its difficulties for estimating $\beta_0$ by an excellent estimation of $\beta_1$, especially with respect to \textit{Omni}. There is little intuition to explain this particular behavior in this more complicated DGP, although 
part of the explanation could probably be related to the general arguments concerning the redistribution-of-mass pointed out by \citet{WWR14} in their Supplementary Materials to which the interested reader is referred. Additionally, similar ambiguous results for the latter technique were observed when considering an alternative estimator for $F_{T|\bX}$, for instance using survival trees as proposed in \citeauthor{WWR14} and using the R code available on their website. In opposition to these results, NEW here satisfactorily mimics the behavior of the omniscient procedure. Simulations for larger samples sizes reveal the same patterns here, and are therefore omitted.

Next, to grasp the impact of a conditional censoring distribution, Table \ref{table:Ex2} highlights the results of DGP \ref{DGP:3} for both sample sizes as well as both censoring percentages. For the sake of brevity, we only report here the case $\tau=0.5$. Similar findings to those previously-described still apply here, as the results of NEW still convincingly mimic the benchmark omniscient estimator in comparison to RM and ICP, especially when considering difficult estimation scenarios. These considerations are naturally reflected when it comes to prediction, as NEW confidently outperforms in this setting both considered weighting schemes.

\begin{table}[t] 
\small
\centering
\begin{tabular}{ccc||C{1cm}C{1cm}C{1cm}C{1cm}C{1cm}C{1cm}|C{1.1cm}}
\multicolumn{3}{c}{$$}&\multicolumn{2}{c}{Bias}&\multicolumn{2}{c}{RMSE}&\multicolumn{2}{c}{MAE}\\
$n$ & $p_c$ & Method & $\beta_0$ & $\beta_1$ & $\beta_0$ & $\beta_1$ &$\beta_0$ & $\beta_1$&MAD\\
\hline
\hline 
\multirow{8}{*}{200} & \multirow{4}{*}{30\%} & \textit{Omni} & 0.000 & 0.007 & 0.259 & 0.374 & 0.168 & 0.240 & 0.330 \\ [1.1pt]
  &&NEW   & -0.033 & 0.024 & 0.272 & 0.374 & 0.175 & 0.258 & 0.341 \\ [1.1pt]
  &&RM    & -0.184 & 0.048 & 0.307 & 0.337 & 0.210 & 0.225 & 0.349 \\ [1.1pt]
  &&ICP   & -0.202 & 0.171 & 0.358 & 0.514 & 0.253 & 0.339 & 0.459 \\ [1.1pt]\cline{3-10}			
											  & \multirow{4}{*}{60\%}& \textit{Omni} & 0.000 & 0.007 & 0.259 & 0.374 & 0.168 & 0.240 & 0.330 \\ [1.1pt]
  &&NEW   & -0.178 & 0.092 & 0.402 & 0.480 & 0.255 & 0.278 & 0.442 \\ [1.1pt]
  &&RM    & -0.518 & 0.151 & 0.584 & 0.381 & 0.521 & 0.278 & 0.584 \\ [1.1pt]
  &&ICP   & -0.899 & 0.339 & 0.998 & 0.694 & 0.951 & 0.451 & 1.024 \\ [1.1pt]\cline{3-10}
\hline
\hline
\multirow{8}{*}{500} & \multirow{4}{*}{30\%} & \textit{Omni} & -0.001 & 0.007 & 0.162 & 0.239 & 0.111 & 0.166 & 0.210 \\ [1.1pt]
  &&NEW   & -0.012 & 0.022 & 0.170 & 0.245 & 0.111 & 0.176 & 0.218 \\ [1.1pt]
  &&RM    & -0.219 & 0.051 & 0.268 & 0.210 & 0.214 & 0.145 & 0.272 \\ [1.1pt]
  &&ICP   & -0.181 & 0.147 & 0.264 & 0.345 & 0.200 & 0.242 & 0.322 \\    [1.1pt]\cline{3-10}			
											  & \multirow{4}{*}{60\%}& \textit{Omni} & -0.001 & 0.007 & 0.162 & 0.239 & 0.111 & 0.166 & 0.210 \\ [1.1pt]
  &&NEW   & -0.077 & 0.037 & 0.232 & 0.298 & 0.158 & 0.196 & 0.278 \\ [1.1pt]
  &&RM    & -0.456 & 0.127 & 0.491 & 0.274 & 0.450 & 0.179 & 0.483 \\ [1.1pt]
  &&ICP   & -0.858 & 0.264 & 0.907 & 0.492 & 0.876 & 0.360 & 0.906 \\[1.1pt]\cline{3-10}
\hline
\hline
\end{tabular}
\caption{\small Simulation results for DGP \DGPref{DGP:3} for the estimation of the median with sample size $n\in\{200,500\}$ and average censoring proportions $p_c \in \{30\%,60\%\}$. Results are expressed in terms of bias, RMSE, MAE and MAD averaged over $B=500$ repetitions. 
}
\label{table:Ex2}
\end{table}

Moving to multivariate covariates, Table \ref{table:Ex3} reports the results of DGP \ref{DGP:4} where four covariates are considered for $\tau=0.5$. For ease of presentation, the results of this DGP are aggregated with respect to the parameters. More specifically, the reported bias stands here for the sum over all parameters of bias taken in absolute value, RMSE stands for the root of mean squared errors summed over all parameters, and MAD stands for the sum over all parameters of median absolute error. Note that in this DGP, and similarly to DGP \ref{DGP:1}, \citeauthor{P03}'s or \citeauthor{PH08}'s estimators would a priori be the most appropriate candidates for estimating $\beta_\tau$ given their inherent global linear assumption. However, as already mentioned earlier, we are mainly interested here in comparing the proposed procedure with the two weighting techniques that have engendered further research in alternative models. One such model is the multivariate example of variable selection in linear regression for which ICP and RM were adapted in \citet{SLZ10} and \citet{WZL13}, respectively. 

As can be observed, the obtained results exhibit analogous patterns to those of the univariate settings; NEW takes advantage of superior bias results to outperform its competitors especially when the proportion of censored responses is important. As a consequence, the predictive potential of NEW also outperforms its competitors in this multivariate setting. 
Finally, it is worth stressing out that for this DGP, the estimator of \citeauthor{WWR14} may suppositionally appear more appropriate to embody the technique of redistribution-of-mass than \citeauthor{WW09}'s estimator, as the former was proposed to avoid kernel smoothing of $F_{T|\bX}$ when confronted to multivariate covariates. However, despite using the available code and tuning the parameters with the recommendations of the paper, the results of \citeauthor{WWR14}'s estimator were here subject to numerical instabilities when studying high censoring proportions, hereby preventing an objective comparison of the methodologies for the considered DGP. Additionally, for lower censoring proportions, the obtained results were analogous to those of RM, and are therefore omitted here in Table \ref{table:Ex3}. 

\begin{table}[t!] 
\small
\centering
\begin{tabular}{cc||ccc|c||ccc|c}
\multicolumn{2}{c}{$$}& \multicolumn{4}{c}{$n=200$}&\multicolumn{4}{c}{$n=500$}\\
$p_c$ & Method & Bias & RMSE & MAE & MAD & Bias & RMSE & MAE & MAD \\
\hline
\hline 
\multirow{4}{*}{30\%} & \textit{Omni} & 0.018 & 0.213 & 0.320 & 0.177 & 0.014 & 0.135 & 0.202 & 0.112 \\ [1.1pt]
  &NEW & 0.062 & 0.270 & 0.405 & 0.220 & 0.140 & 0.184 & 0.281 & 0.150\\ [1.1pt]
  &RM  & 0.192 & 0.268 & 0.409 & 0.219 & 0.257 & 0.206 & 0.337 & 0.167 \\ [1.1pt]
  &ICP & 0.271 & 0.352 & 0.547 & 0.283 & 0.252 & 0.248 & 0.379 & 0.199 \\ [1.1pt]\cline{2-10}			
\multirow{4}{*}{60\%}& \textit{Omni} & 0.018 & 0.213 & 0.320 & 0.177 & 0.014 & 0.135 & 0.202 & 0.112 \\ [1.1pt]
  &NEW & 0.224 & 0.394 & 0.595 & 0.304 & 0.313 & 0.293 & 0.457 & 0.237 \\ [1.1pt]
  &RM  & 0.581 & 0.448 & 0.731 & 0.364 & 0.559 & 0.364 & 0.608 & 0.296 \\ [1.1pt]
  &ICP & 1.044 & 0.716 & 1.153 & 0.560 & 1.017 & 0.621 & 1.096 & 0.493 \\ [1.1pt]
\hline
\hline
\end{tabular}
\caption{\small Simulation results for DGP \DGPref{DGP:4} for the estimation of the median. Results are in terms of aggregated over the parameters bias, RMSE, MAE and MAD averaged over $B=500$ repetitions. 
}
\label{table:Ex3}
\end{table}
Lastly, to evaluate the effectiveness of the percentile bootstrap described in Section \ref{section:prop} for the proposed procedure, we compare the performance of the latter with the percentile bootstrap of RM for DGPs \DGPref{DGP:1} and \DGPref{DGP:2}. Table \ref{table:Bootstrap} reports the empirical coverage probability and empirical mean length of the confidence intervals obtained with 300 bootstrapped samples at each of 500 iterations. The nominal level is taken to be 0.95, and for both procedures the bandwidths are fixed at 0.05 for DGP \DGPref{DGP:1} and 0.10 for DGP \DGPref{DGP:2}. As can first be observed, for the simplest DGP \DGPref{DGP:1} both procedures yield coverage probabilities adequately close to the chosen nominal level and with analogous mean empirical length of the confidence intervals. Secondly, for a more complicated scenario with namely more censoring, we observe here that NEW still appropriately approaches the nominal level while RM exhibits difficulties for the latter. We note that this could partly be due to an inappropriate choice of bandwidth. Nevertheless, these results suggest here that the percentile bootstrap represents a satisfactory tool for inference purposes when considering the estimator NEW.

To conclude, from the described simulation results we observe that NEW provides a valuable complement to the literature, especially in cases where the latter seem to fail when confronted to high censoring percentages and quantile levels. Moreover, for the simpler setting considered here, NEW offers similar results to those of RM which are often taken as primary reference. These considerations are believed to be encouraging for the practical use of the newly proposed loss function for quantile regression models. 

\begin{table}[t!] 
\small
\centering
\begin{tabular}{ccc||cccc|cccc}
\multicolumn{3}{c}{$$}&\multicolumn{4}{c}{ECP}&\multicolumn{4}{c}{EML}\\
\multicolumn{3}{c}{$$}&\multicolumn{2}{c}{RM}&\multicolumn{2}{c}{NEW}&\multicolumn{2}{c}{RM}&\multicolumn{2}{c}{NEW}\\
DGP & $p_c$ & $n$ & $\beta_0$ & $\beta_1$ & $\beta_0$ & $\beta_1$ &$\beta_0$ & $\beta_1$&$\beta_0$ & $\beta_1$\\
\hline
\hline 
\multirow{4}{*}{1 ($\tau=0.5$)} & \multirow{2}{*}{15\%} & 100 &  0.954 & 0.952 & 0.948 & 0.952 & 1.049  & 1.894 & 1.052 & 1.902\\ [1.1pt]
  &&200 & 0.936& 0.950 & 0.938 & 0.960 & 0.752 & 1.342& 0.752& 1.346\\ [1.1pt] \cline{3-11}
  & \multirow{2}{*}{40\%} & 100 &0.956 & 0.958 &0.966 &0.970  & 1.162 & 2.258 &1.187 & 2.292\\ [1.1pt]
  &&200 & 0.940 & 0.952 & 0.940 & 0.954 & 0.839 & 1.587 & 0.837 & 1.589\\ [1.1pt]
\hline
\hline		
\multirow{4}{*}{2 ($\tau=0.3$)} & \multirow{2}{*}{30\%} & 100 & 0.956 & 0.956 & 0.960 & 0.948 & 2.085 & 2.692 & 2.218 & 2.835 \\ [1.1pt]
  &&200 & 0.922 & 0.946 & 0.972& 0.954 & 1.490 & 1.872 & 1.543& 1.990\\ [1.1pt] \cline{3-11}
  & \multirow{2}{*}{60\%} & 100 &0.758 & 0.944 & 0.968 & 0.942  & 1.864 & 2.315 &2.301 &2.892 \\ [1.1pt]
  &&200 & 0.756 & 0.950 & 0.960 & 0.966 & 1.463 & 1.716& 1.609 & 1.970 \\ [1.1pt] 
\hline
\hline
\end{tabular}
\caption{\small Bootstrap results for DGP \DGPref{DGP:1} and \DGPref{DGP:2} based on 500 simulations with $300$ bootstrap samples. ECP and EML stand respectively for empirical coverage probability and empirical mean length, for a nominal level of 0.95. Average censoring proportions $p_c$ are taken in $\{15\%,30\%,40\%,60\%\}$, and quantile levels $\tau$ are chosen among $\{0.3,0.5\}$.
}
\label{table:Bootstrap}
\end{table}

\section{Real Data Analysis}\label{section:application}
As an illustration, we propose to apply the developed methodology to the Channing House dataset which is readily available in the R package \texttt{boot}. The data consists of 462 records of residents living at the retirement center `Channing House' during the period January 1964 to July 1975. One of the purposes of the study was to assess the difference between the survival of men and women under the retirement center program, taking age of entry into account. To that end, the dataset contains information about the survival time (in months) of individuals, their sex  and their age (in months) at the time they entered the retirement center. At the end of the study, only 176 residents taken into account experienced the event of interest, resulting in approximately 62\% of censoring. Further information on the dataset may be found for instance in \citet{H80}.

Embracing the role of a practitioner, as a preliminary remark in the objective of fitting a quantile regression model to the data, we note that the responses are subject to a considerable amount of censoring, hereby ruling out a confident use of inverse-censoring-probability-based methodologies, and in particular \citeauthor{BT02}' estimator as illustrated in our simulations. Therefore, we will only consider for this analysis \citeauthor{WW09}'s estimator as a point of comparison for the proposed methodology.
     
In order to examine the effects of the covariates on several quantile levels of the survival time, we now consider first a sequence of quantiles equally ranging from 0.1 to 0.5 with 0.05 increments. For every quantile level, we estimate $\beta_\tau$ with both \citeauthor{WW09}'s estimator and the proposed methodology. Implementation of the latter is carried out using the algorithm of Section \ref{section:minalgo}, where, as in Section \ref{section:simulation}, we adopt Beran's estimator on each sex for $\widehat{G}_C$. \citeauthor{WW09}'s estimator is likewise implemented with Beran's estimator on each sex for $\widehat{F}_{T|\bX}$. Concerning the bandwidths of both estimators, equivalently to our simulation study, the latter are chosen using 5-fold cross validation on the normalized covariate age among 15 candidates equally ranging from 0.05 to 1.5. Finally, for both procedures, 95\% confidence intervals are constructed using the percentile bootstrap described in Section \ref{section:prop} based on 300 bootstrap samples. For computational convenience, all bootstrap estimations for both methodologies are based on the same bandwidths as for the initial dataset.

\begin{figure}[t!]
		 \captionsetup[subfigure]{labelformat=empty}
  	 \centering
     \subfloat[][]{\includegraphics[scale=.335]{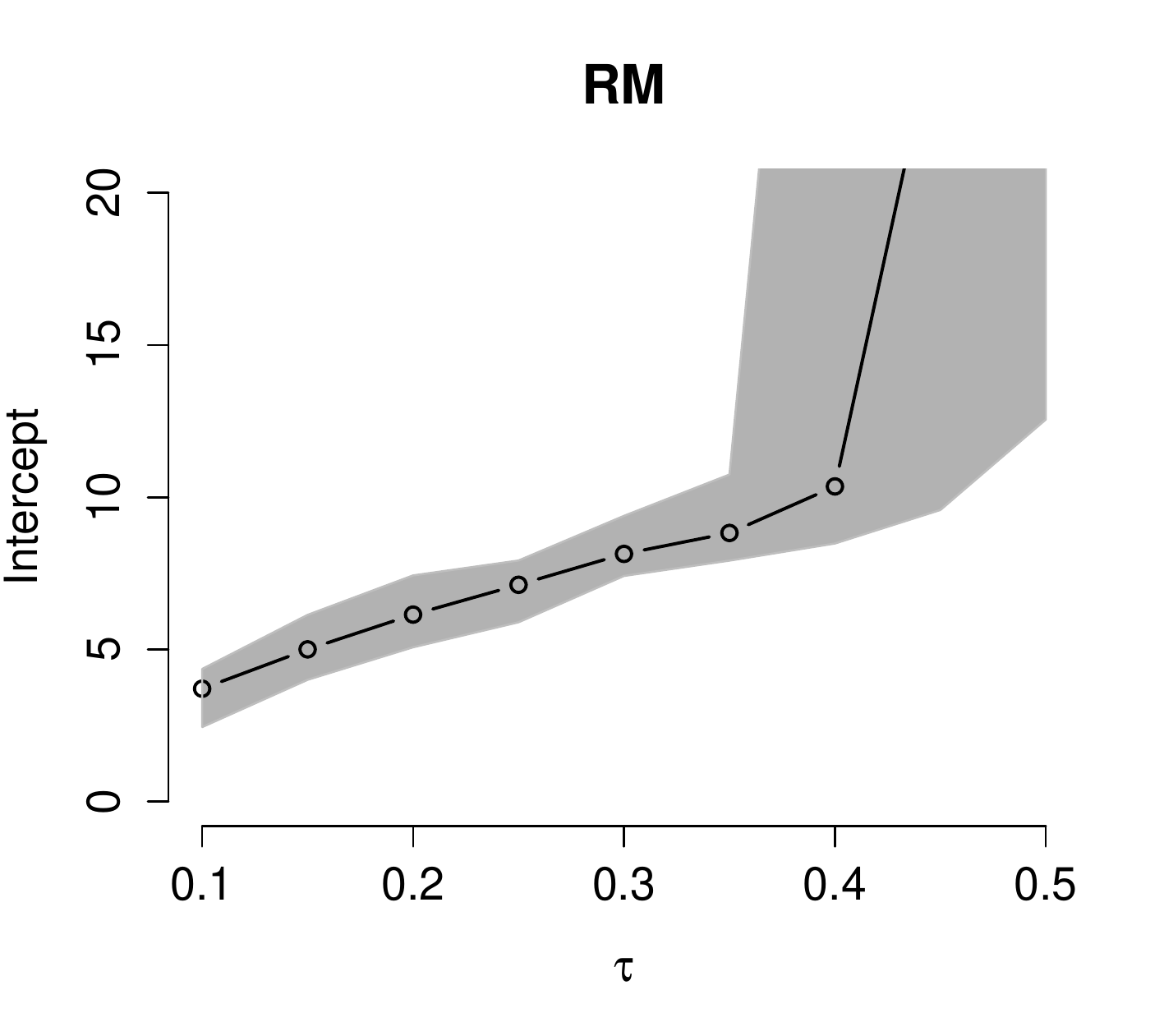}}
     \hspace{0.5cm}
     \subfloat[][]{\includegraphics[scale=.335]{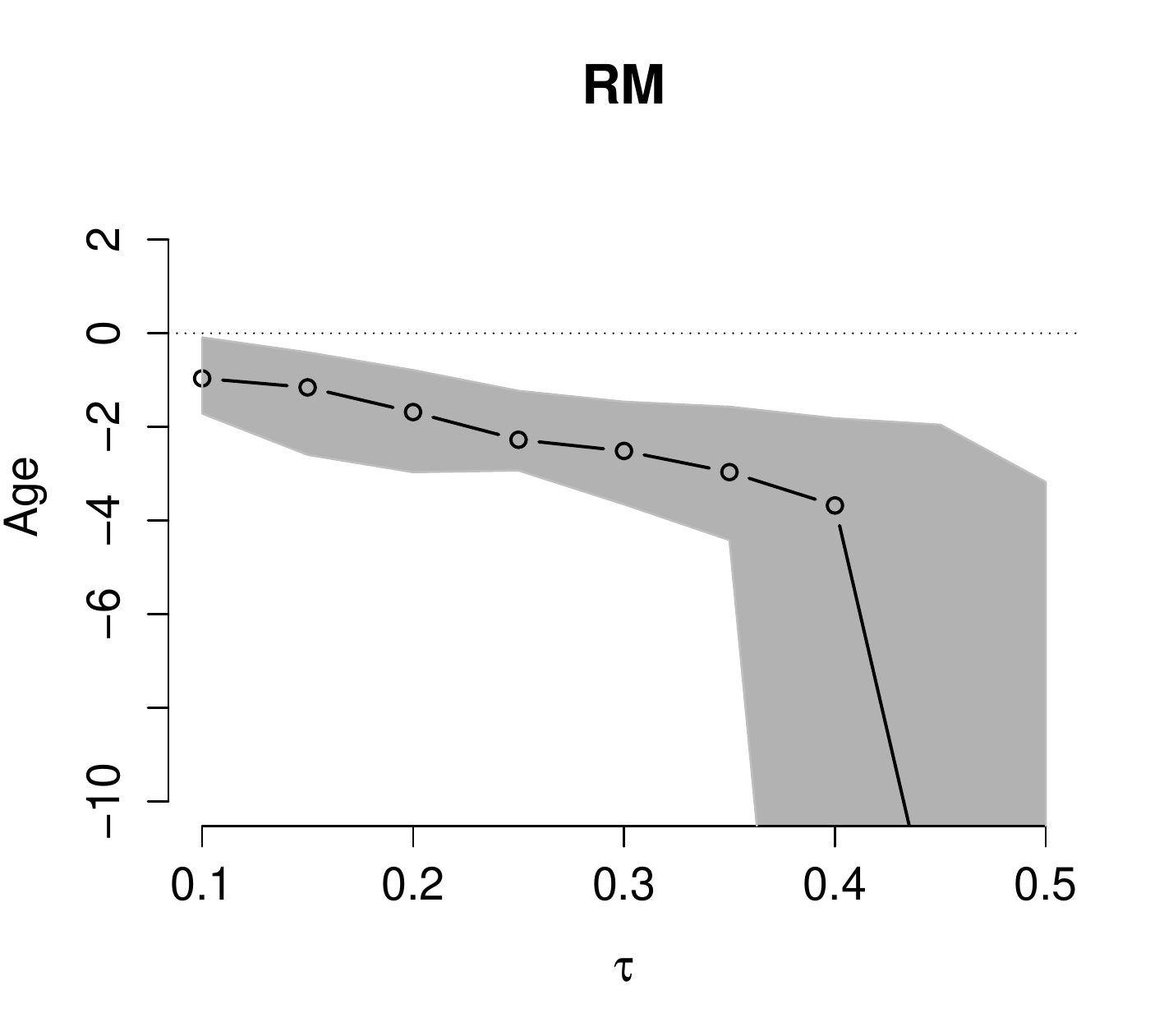}}
     \hspace{0.5cm}
     \subfloat[][]{\includegraphics[scale=.335]{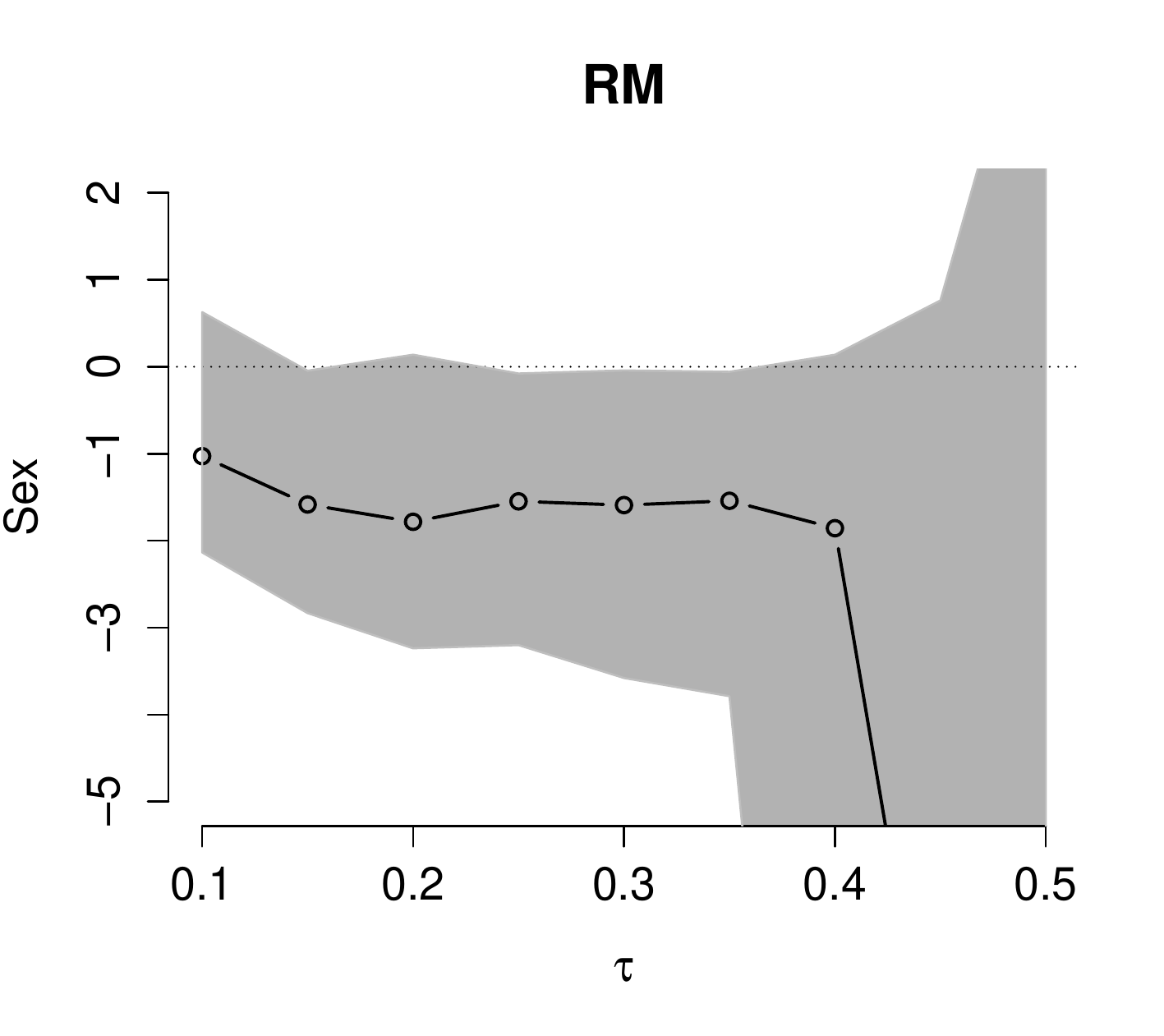}}
     \hspace{0.5cm}
     \subfloat[][]{\includegraphics[scale=.335]{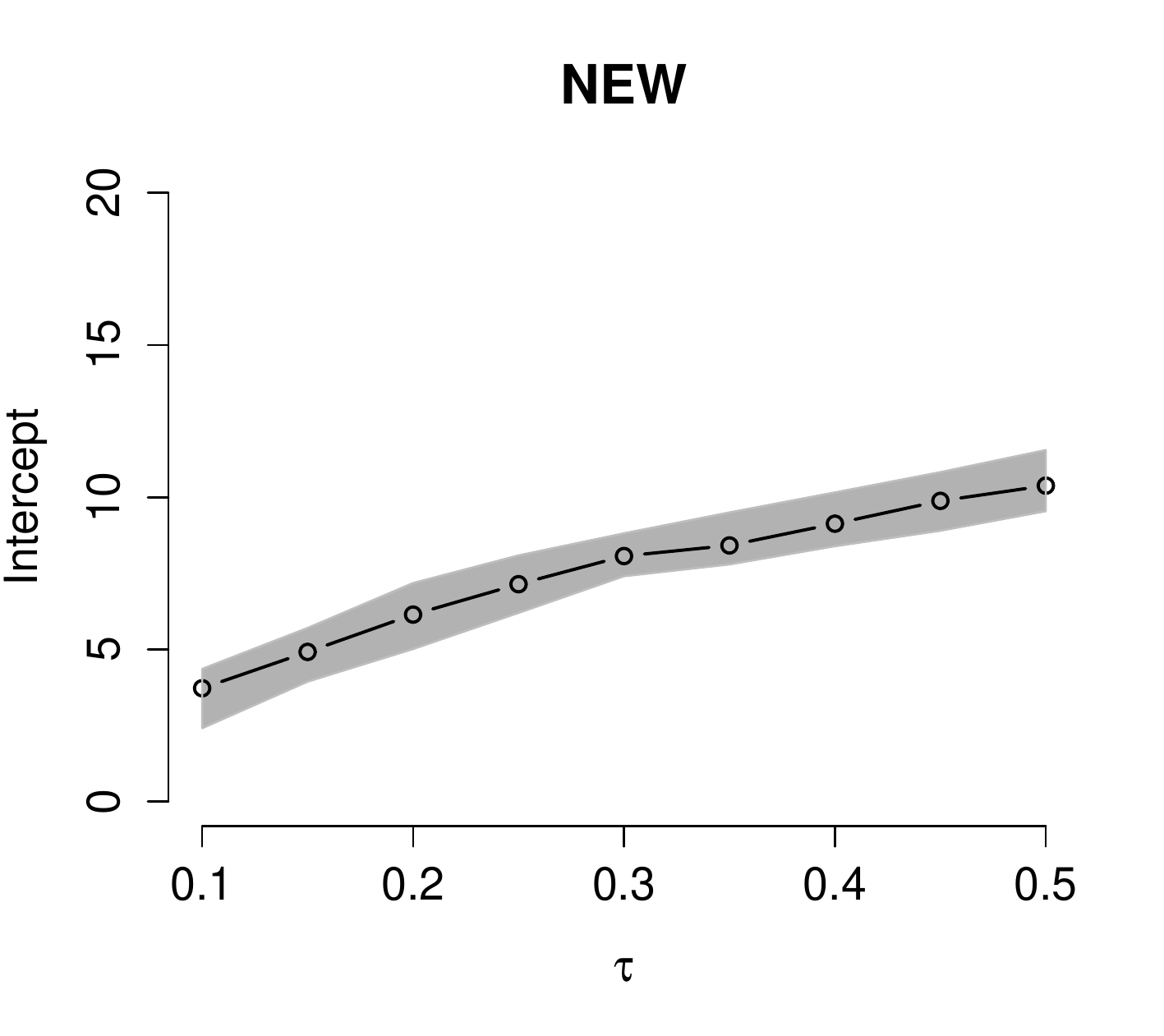}}
     \hspace{0.5cm}
     \subfloat[][]{\includegraphics[scale=.335]{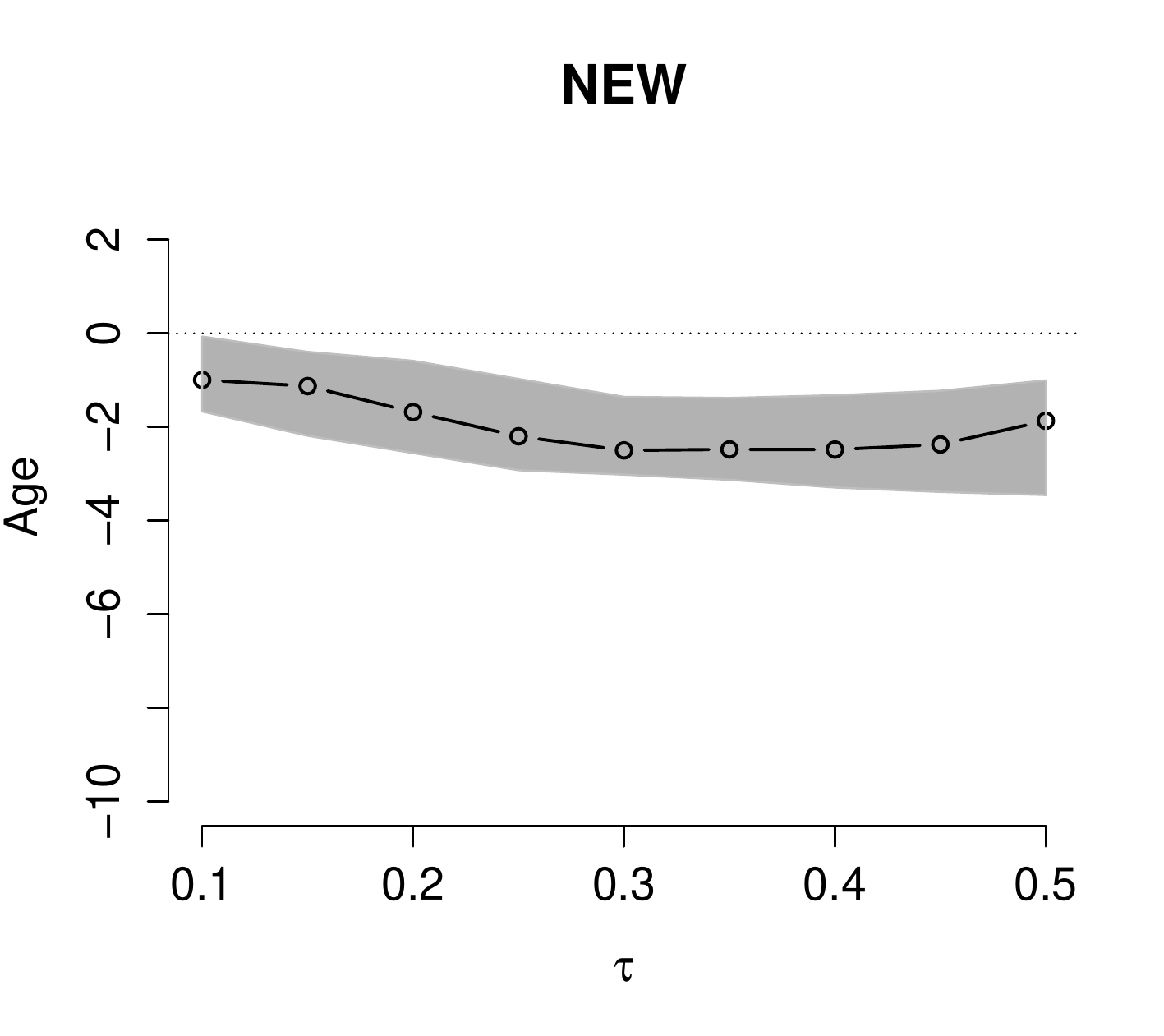}}
     \hspace{0.5cm}
     \subfloat[][]{\includegraphics[scale=.335]{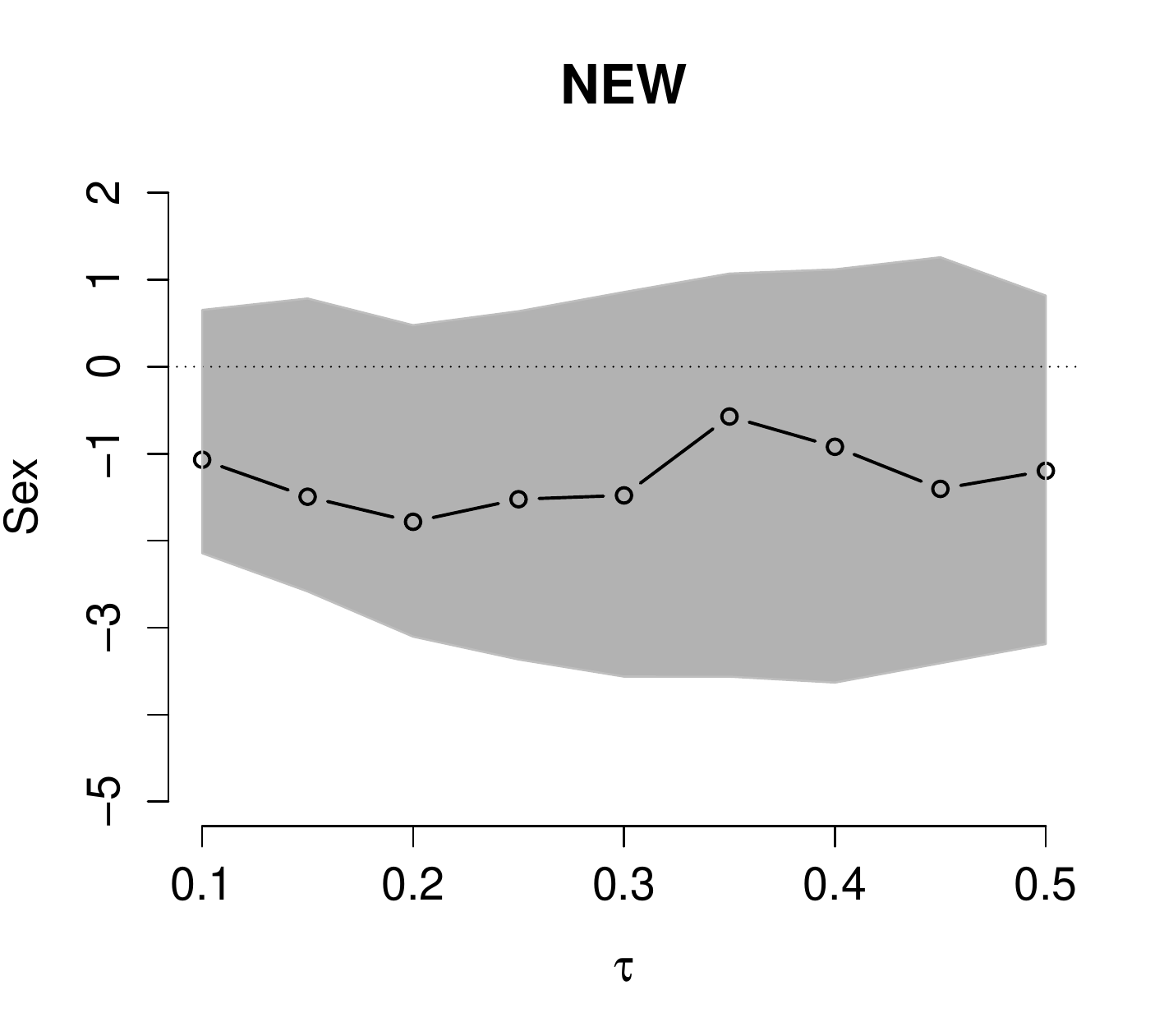}}
		\caption{\small Channing House data. Estimated quantile coefficients for regressing the survival time (in years) on sex and age at entry (standardized). The top row reports the results for \citeauthor{WW09}'s estimator (RM), while the bottom row depicts results obtained with the present methodology (NEW). Shaded areas correspond to 95\% confidence intervals based on 300 bootstrap samples.}
     \label{fig:App_beta}
\end{figure}

Estimation results are presented in Figure \ref{fig:App_beta}. A first consideration one may acknowledge from the presented figure is the noticeable numerical stability offered by the proposed methodology in comparison to \citeauthor{WW09}'s estimator when estimating relatively high quantile levels with respect to the censoring proportion. This appears as a natural consequence of requiring for our procedure the estimation of $G_C$ rather than $F_{T|\bX}$, hereby highlighting again the potential complement our estimator may provide to the literature for relatively highly censored datasets. Apart from this first examination, considering only quantile levels below the region of numerical instability of RM, both methods suggest unsurprisingly a significant negative effect of age at entry on the survival time, while there seems to be a discrepancy on the significance of a gender effect for the present dataset even though the upper confidence band of RM is very close to fully incorporating the value 0. Lastly, examining the bootstrap confidence intervals, we observe that confidence intervals for NEW tend to be a little smaller than for RM except for the covariate sex. Furthermore, it is worth stressing out that the width of the NEW's confidence intervals does not seem here to fluctuate much with the increasing quantile levels of interest, hereby exposing a confident resistance to the risk of estimating higher quantiles with a consequent proportion of censored responses.

To further examine the performance of both procedures for the present dataset, we consider a prediction study based on cross validation. Specifically, we first randomly split the data into a training set of 350 observations and a testing set of 112 observations. For quantile levels ranging only from 0.1 to 0.4 by 0.05 increments given the instability of RM beyond $\tau=0.4$, we then estimate the quantile coefficients for each method based on the training set, and consequently evaluate the predictive performance of the $\tau-$th conditional quantile of fully observed responses in the testing set. As an evaluation measure, we therefore consider the following prediction error: 
$$\text{PE} =  \text{med}_{\stackrel{i\in \text{testing}}{\Delta_i=1}}\; \rho_\tau\left(Y_i - \widehat{\beta}_\tau^{\mathsf{T}} \bX_i \right),$$ where $\widehat{\beta}_\tau$ is an estimator of $\beta_\tau$ based on the training set, and $Y_i, i=1,\ldots,n$, denote the observed survival times. For robust results, this cross validation is repeated 200 times and the median of all PE's is reported in Table \ref{table:App_PE}. Similarly to the above-described bootstrap confidence intervals, for computational consideration, the bandwidths of both NEW and RM are selected by cross validation on one initial training set. Results suggest that for this dataset, the prediction accuracies of both procedures are very similar, with a slight overall advantage for NEW. Given the further improvement of NEW over RM for quantile levels lying above $\tau=0.4$, this observation may then advocate here for the practical choice of the proposed methodology for a robust quantile regression analysis. 

\begin{table}[t!] 
\footnotesize
\centering
\begin{tabular}{c||ccccccc}
Method & $\tau=0.1$ & $\tau=0.15$ & $\tau=0.2$ & $\tau=0.25$ & $\tau=0.3$ & $\tau=0.35$ &$\tau=0.4$ \\
\hline
\hline
&&&&&&&\\[-7pt]
RM  & 0.474  & 0.647  & 0.876  & 1.112  & 1.225  & 1.522  & 2.103 \\[-1pt]
    & (0.02) & (0.04) & (0.07) & (0.10) & (0.10) & (0.11) & (0.10)\\[1.2pt] 
NEW & 0.474  & 0.650  & 0.853  & 1.065  & 1.225  & 1.437  & 1.774 \\[-1pt]
    & (0.02) & (0.03) & (0.05) & (0.05) & (0.11) & (0.11) & (0.10)\\[1.2pt] 
\hline
\hline
\end{tabular}
\caption{\small Channing House data. Median prediction error over 200 cross validations, with standard deviations of the latter in parentheses, for quantile levels $\tau \in \{0.1,\ldots,0.4\}$ for both RM and NEW.}
\label{table:App_PE}
\end{table}

\section{Conclusion}
In this work we have proposed an adapted version of the check function for evaluating sample quantiles or more generally quantile regression when confronted to possible right-censored responses. By tackling the problem of incomplete data at the level of the loss function through an adjustment of the underlying penalization for under- and overestimation of the true quantile value, the proposed loss function allows to plainly take profit of all observations at hand. This has to be contrasted for instance to the prominent inverse-censoring-probability weighting scheme of \citet{KSVR81}. 

As an illustration of the newly proposed loss function for regression models, we proposed to consider the particular and well-studied case of a simple linear regression, for which consistency of the quantile coefficient estimator was obtained using recent results on non-smooth semiparametric estimation equations with an infinite-dimensional nuisance parameter. For practical implementation, a detailed adaptation of the MM algorithm was then proposed, and an extensive simulation study was carried out to illustrate the finite sample performance of the methodology. From the latter study, the proposed estimator was observed to perform competitively with respect to the established technique of redistribution-of-mass, especially when considering highly censored datasets. This consideration was highlighted in a real data application as well. Furthermore, for simpler settings, the proposed estimator revealed to be interestingly robust to both the sample size and the smoothing parameter to be selected for a preliminary, and possibly local, estimator of the censoring distribution on which the procedure is built. Lastly, for multivariate settings, while the present paper only considers Beran's estimator for the conditional censoring distribution, we note that alternative modelling techniques for the latter may be more appropriate to avoid the curse of dimensionality, such as a Cox model, a single-index model or survival trees for instance. Such developments deserve further analysis.

To conclude, considering now a broader picture than strictly parametric regression, the proposed results also illustrate the potential of the studied loss function for improvement of various modelling techniques that are currently built upon the inverse-censoring-probability weights, such as single-index regression or copula-based regression to name a few. This provides encouraging perspectives for enhancement of flexible censored quantile regression models. 


\section*{Acknowledgments}
All authors acknowledge financial support from IAP research network P7/06 of the Belgian Government (Belgian Science Policy). M. De Backer and A. El Ghouch further acknowledge financial support from the FSR project IMAQFSR15PROJEL from the Fonds de la Recherche Scientifique de Belgique (F.R.S.-FNRS). I. Van Keilegom also acknowledges support from the European Research Council (2016-2021, Horizon 2020 / ERC grant agreement No.\ 694409).

Computational resources have been provided by the supercomputing facilities of the Universit\'{e} catholique de Louvain (CISM/UCL) and the Consortium des \'{E}quipements de Calcul Intensif en F\'{e}d\'{e}ration Wallonie Bruxelles (C\'{E}CI) funded by the Fonds de la Recherche Scientifique de Belgique under convention 2.5020.11.

\newpage
\section*{Appendix}\label{section:appendix} 
We provide in this appendix the proof of consistency of the proposed estimator, which relies heavily on the work of \citet{DVK15} (DVK hereafter) on non-smooth semiparametric $M$-estimation problems. As a preliminary remark, and as already mentioned in Section \ref{section:prop}, note that the following proof is built on a crucial result of \citet{L11} for the class $\mathcal{G}$ in \ref{cond:True G_C deriv}. This implies that our proof has to be read as such under similar conditions as for \citeauthor{L11}, that is, assuming the existence of a function $g: \RR^{d+1} \rightarrow \RR$ such that $G_C(\cdot|\bX)=G_C(\cdot|g(\bX))$, or simply considering the case of a univariate covariate. For ease of reading, the proof is written considering the latter case.
\renewcommand{\theequation}{A.\arabic{equation}}
\begin{proof}[\nopunct] \textbf{Proof of Theorem \ref{theorem consist}}. In Theorem 1 of DVK, five high level conditions (A1)-(A5) are developed under which an $M$-estimator is consistent in a general semiparametric maximization problem. We therefore only need to verify the latter conditions to prove the consistency of $\widehat{\beta}_\tau$. For notational convenience with respect to the work of DVK, let us first rewrite $\widehat{\beta}_\tau$ as  
\begin{eqnarray}\label{eq:Cens_esti_max}
\widehat{\beta}_\tau = \arg\max_{\beta \in \mathcal{B}}\sum_{i=1}^n \psit\left(\beta^{\mathsf{T}} \bX_i;Y_i,\widehat{G}_C(\cdot|\bX_i)\right),
\end{eqnarray}
where $\psit(a;y,G)=-\rhot(a;y)+(1-\tau)\int_0^{a}G(s) \, \mathrm ds$ and where $\mathcal{B}$ is a compact parameter space, taken to be the neighborhood of $\beta_\tau$ mentioned in our assumptions. For further convenience and without any loss of generality, the proof is written considering the response variable $Y$ to be positive. This consideration is solely done in the purpose of being coherent with the arbitrary set value of 0 in the correcting term of $\psit$ with respect to $v$ defined in assumption \ref{cond:True G_C deriv}. For instance, one could also easily consider in the following a strictly negative variable $Y$, but this would possibly require the arbitrary chosen constant 0 to be replaced by any constant below $v$ as we only wish to control the behavior of the nuisance parameter $\widehat{G}_C(\cdot|\bX)$ below this $v$.

Next, starting with condition (A1) in DVK, note that the latter is readily satisfied in our framework by construction of $\widehat{\beta}_\tau$. Furthermore, using the definition of $\mathcal{G}$ in \ref{cond:True G_C deriv} as the space embedding the nuisance parameter $G_C$, and equiping the latter with the distance $d_{\mathcal{G}}(G_1,G_2)=\sup_{\bx \in \text{supp}(\bX)}\sup_{y \leq \upsilon}\left|G_1(y|\bx)-G_2(y|\bx)\right|$ for any $G_1,G_2 \in \mathcal{G}$, note that (A3) in DVK is straightforwardly satisfied as well provided  assumption \ref{cond:Ghat} holds. We therefore only need to verify here conditions (A2), (A4) and (A5). 

Starting with the identifiability condition (A2) ensuring the uniqueness of $\beta_\tau$, we need to verify that for any $\epsilon > 0$, $\inf_{||\beta-\beta_\tau||>\epsilon} \E\left[\psit\left(\beta^{\mathsf{T}}_\tau \bX;Y,G_C(\cdot|\bX)\right)-\psit\left(\beta^{\mathsf{T}} \bX;Y,G_C(\cdot|\bX)\right)\right]>0$, where $||\cdot||$ denotes the Euclidean distance. To that end, using the definition of $\vt$, we can show that
\begin{align*} 
\begin{split}
&\phantom{=}\; \inf_{||\beta-\beta_\tau||>\epsilon}\E\left[\psit\left(\beta^{\mathsf{T}}_\tau \bX;Y,G_C(\cdot|\bX)\right)-\psit\left(\beta^{\mathsf{T}} \bX;Y,G_C(\cdot|\bX)\right)\right] \\
&=\inf_{||\beta-\beta_\tau||>\epsilon}\E\left[\int_{\beta^{\mathsf{T}} \bX}^{\beta^{\mathsf{T}}_\tau \bX}\left(\ind(Y \geq s)-(1-\tau)\bar{G}_C(s|\bX)\right)\mathrm ds\right]\\
&=\inf_{||\beta-\beta_\tau||>\epsilon}\E\left[\int_{\beta^{\mathsf{T}} \bX}^{\beta^{\mathsf{T}}_\tau \bX}(1-G_C(s|\bX))(\tau-F_{T|\bX}(s|\bX))\mathrm ds\right],
\end{split} 
\end{align*}
where the latter expectation taken with respect to the distribution of $\bX$ is strictly positive under assumptions \ref{cond:supp X}, \ref{cond:true cond dens} and \ref{cond:G_C}, hereby ensuring condition (A2) is satisfied.

Next, for (A4) to hold, it suffices by Remark 1(ii) in DVK and assumption \ref{cond:True G_C deriv} to show that the class $$\mathcal{F}=\left\{(y,\bx) \mapsto \psit\left(\beta^{\mathsf{T}} \bx;y,G(\cdot|\bx)\right): \beta \in \mathcal{B}, G \in \mathcal{G} \right\}$$ is Glivenko-Cantelli. For this, by Theorem 2.4.1 in \citet{VdVW96}, we need to prove that for all $\epsilon > 0$, the $\epsilon-$bracketing number $N_{[\,]}(\epsilon, \mathcal{F}, L_1(P))$ of the class $\mathcal{F}$ with respect to the $L_1$ probability measure on $(Y,\bX)$ is finite. To that end, let $\psit = \psi_{\tau 1} + \psi_{\tau 2} + \psi_{\tau 3},$ where
 \begin{align*}
	\psi_{\tau 1}\left(\beta^{\mathsf{T}} \bx;y,G(\cdot|\bx)\right) &= -\tau(y-\beta^{\mathsf{T}} \bx)\\ 
	\psi_{\tau 2}\left(\beta^{\mathsf{T}} \bx;y,G(\cdot|\bx)\right) &= (y-\beta^{\mathsf{T}} \bx)\ind(y<\beta^{\mathsf{T}} \bx)\\
	\psi_{\tau 3}\left(\beta^{\mathsf{T}} \bx;y,G(\cdot|\bx)\right) &= (1-\tau)\int_0^{\beta^{\mathsf{T}} \bx} G(s|\bx)\,\mathrm ds,
\end{align*}
and let $\mathcal{F}_1, \mathcal{F}_2$ and $\mathcal{F}_3$ denote the classes induced by $\psi_{\tau 1}, \psi_{\tau 2}$ and $\psi_{\tau 3}$, respectively. From this decomposition, it is easy to see that 
\begin{eqnarray}\label{eq:bracket_prod}
N_{[\,]}(\epsilon, \mathcal{F}, L_1(P))\leq \prod_{j=1}^3 N_{[\,]}(\epsilon, \mathcal{F}_j, L_1(P)).
\end{eqnarray}
Now, for the classes $\mathcal{F}_1$ and $\mathcal{F}_2$, suppose for simplicity and without loss of generality that all coordinates of $\bx$ are positive, and define $M_\epsilon = O(\epsilon^{-2})$ pairs $(\beta_k^L,\beta_k^U), k=1,\ldots,M_\epsilon,$ that cover $\mathcal{B}$, assumed to be compact by \ref{cond:supp X}, such that $(\beta_k^{L_{\mathsf{T}}}\bx,\beta_k^{U_{\mathsf{T}}}\bx)$ define brackets of length $\epsilon\tau/(1-\tau)$ for the class $\{\bx \mapsto \beta^{\mathsf{T}} \bx: \beta \in \mathcal{B}\}$ with respect to the $L_1-$norm. Then, it is straightforward that $N_{[\,]}(\epsilon, \mathcal{F}_j, L_1(P)) \leq K_j\epsilon^{-2}$ for some finite constants $K_j>0, j=1,2,$ which, combined with \eqref{eq:bracket_prod}, suggests we only have to verify that $N_{[\,]}(\epsilon, \mathcal{F}_3, L_1(P))$ is bounded in order to prove that condition (A4) holds in our framework. 

To that end, by Lemma 6.1. in \citet{L11} which extends Theorem 2.7.5 in \citeauthor{VdVW96}, first note that there exist $N_\epsilon \leq \exp(K_3\epsilon^{-2/(1+\eta)})$ brackets $(\underline{G}_j,\overline{G}_j), j=1,\ldots,N_\epsilon,$ for a finite constant $K_3>0$ such that, under \ref{cond:True G_C deriv}, for all $G \in \mathcal{G}$, there exists $j=1,\ldots,N_\epsilon,$ for which $\underline{G}_j\leq G\leq\overline{G}_j$, and 
\begin{eqnarray}\label{eq:Lemma_Lopez}
\int_{\text{supp}(\bX)}\int_0^{\upsilon} \left|\overline{G}_j(s|\bx)-\underline{G}_j(s|\bx) \right| \mathrm ds \, \mathrm dF_\bX(\bx) < \epsilon,
\end{eqnarray}
where $F_\bX(\bx)$ denotes the c.d.f. of $\bX$. From this result, our claim for (A4) to hold is that brackets for $\mathcal{F}_3$ are given by $(\underline{\zeta}_{jk},\overline{\zeta}_{jk}), j=1,\ldots,N_\epsilon$, $k=1,\ldots,M_\epsilon,$ where 
\begin{align*}
\underline{\zeta}_{jk}(\bx) &= (1-\tau)\int_0^{\beta_k^{L_{\mathsf{T}}}\bx}\underline{G}_j(s|\bx)\,\mathrm ds,\\
\overline{\zeta}_{jk}(\bx)  &= (1-\tau)\int_0^{\beta_k^{U_{\mathsf{T}}}\bx}\overline{G}_j(s|\bx)\,\mathrm ds. 
\end{align*}
For this claim to hold, as it is straightforward to verify that for all $\zeta \in \mathcal{F}_3$ there exist $j=1,\ldots,N_\epsilon,$ and $k=1,\ldots,M_\epsilon$, such that $\underline{\zeta}_{jk}\leq \zeta \leq\overline{\zeta}_{jk}$, we only need to show that 
$$\int_{\text{supp}(\bX)} \left|\overline{\zeta}_{jk}(\bx)-\underline{\zeta}_{jk}(\bx) \right| \mathrm dF_\bX(\bx) < \epsilon, \quad j=1,\ldots,N_\epsilon,\; k=1,\ldots,M_\epsilon.$$ To that end, developing the expressions of $\underline{\zeta}_{jk}$ and $\overline{\zeta}_{jk}$, we have that 
\begin{align*} 
\begin{split}
&\phantom{=}\; \int_{\text{supp}(\bX)}\left|\overline{\zeta}_{jk}(\bx)-\underline{\zeta}_{jk}(\bx) \right| \mathrm dF_\bX(\bx) \\
&= (1-\tau)\int_{\text{supp}(\bX)} \left|\int_0^{\beta_k^{U_{\mathsf{T}}}\bx}\overline{G}_j(s|\bx)\mathrm ds-\int_0^{\beta_k^{L_{\mathsf{T}}}\bx}\underline{G}_j(s|\bx)\mathrm ds \right| \mathrm dF_\bX(\bx),
\end{split} 
\end{align*}
where the latter expression can be bounded above by $T_1 + T_2$ where
\begin{align*} 
T_1 &= (1-\tau)\int_{\text{supp}(\bX)} \left|\int_0^{\beta_k^{U_{\mathsf{T}}}\bx}\overline{G}_j(s|\bx)\mathrm ds-\int_0^{\beta_k^{U_{\mathsf{T}}}\bx}\underline{G}_j(s|\bx)\mathrm ds \right| \mathrm dF_\bX(\bx),\\
T_2 &= (1-\tau)\int_{\text{supp}(\bX)} \left|\int_0^{\beta_k^{U_{\mathsf{T}}}\bx}\underline{G}_j(s|\bx)\mathrm ds-\int_0^{\beta_k^{L_{\mathsf{T}}}\bx}\underline{G}_j(s|\bx)\mathrm ds \right|\mathrm dF_\bX(\bx).
\end{align*}
We will now show that both $T_1$ and $T_2$ can be bounded above such that their sum is bounded by $\epsilon$. Starting with $T_1$, we have that 
\begin{align*} 
T_1 &\leq (1-\tau)\int_{\text{supp}(\bX)}\int_0^{\beta_k^{U_{\mathsf{T}}}\bx} \left|\overline{G}_j(s|\bx)-\underline{G}_j(s|\bx)\right| \mathrm ds \,\mathrm dF_\bX(\bx)\\
&\leq (1-\tau)\int_{\text{supp}(\bX)}\int_0^{\upsilon} \left|\overline{G}_j(s|\bx)-\underline{G}_j(s|\bx)\right| \mathrm ds\,\mathrm dF_\bX(\bx) \leq (1-\tau)\epsilon, 
\end{align*}
using assumption \ref{cond:G_C} and \eqref{eq:Lemma_Lopez} for the second and last inequalities, respectively. Concentrating now on $T_2$, we have that
\begin{align*} 
T_2 &\leq (1-\tau)\int_{\text{supp}(\bX)} \int_{\beta_k^{L_{\mathsf{T}}}\bx}^{\beta_k^{U_{\mathsf{T}}}\bx} \left|\underline{G}_j(s|\bx)\right| \mathrm ds \, \mathrm dF_\bX(\bx)\\
&\leq (1-\tau)\int_{\text{supp}(\bX)}\left|\beta_k^{U_{\mathsf{T}}}\bx-\beta_k^{L_{\mathsf{T}}}\bx\right|\mathrm dF_\bX(\bx) \leq \tau\epsilon, 
\end{align*}
given the brackets induced by $(\beta_k^{L_{\mathsf{T}}},\beta_k^{U_{\mathsf{T}}})$ for the class $\{\bx \mapsto \beta^{\mathsf{T}} \bx: \beta \in \mathcal{B}\}$ with respect to the $L_1-$norm. This completes the proof that $N_{[\,]}(\epsilon, \mathcal{F}_3, L_1(P))$ is bounded. Hence, we conclude that $N_{[\,]}(\epsilon, \mathcal{F}, L_1(P))=O(\exp(K_3\epsilon^{-2/(1+\eta)}))$, from which it follows that condition (A4) holds.

Lastly, for condition (A5), we need to establish that 
$$
\lim_{d_{\mathcal{G}}(G,G_C)\rightarrow 0} \sup_{\beta \in \mathcal{B}} \left| \E\left[ \psit\left(\beta^{\mathsf{T}} \bX;Y,G(\cdot|\bX)\right) - \psit\left(\beta^{\mathsf{T}} \bX;Y,G_C(\cdot|\bX)\right) \right]\right| = 0.
$$
To that end, note that  
\begin{align*} 
&\phantom{\leq}\; \sup_{\beta \in \mathcal{B}} \left| \E\left[ \psit\left(\beta^{\mathsf{T}} \bX;Y,G(\cdot|\bX)\right) - \psit\left(\beta^{\mathsf{T}} \bX;Y,G_C(\cdot|\bX)\right) \right]\right|\\
&\leq (1-\tau) \sup_{\beta \in \mathcal{B}} \E\left[ \int_0^{\beta^{\mathsf{T}} \bX}\left|G(s|\bX)-G_C(s|\bX)\right|\mathrm ds \right],
\end{align*}
where the last expectation is taken with respect to the distribution of $\bX$. Under assumption \ref{cond:G_C}, this expression can then in turn be bounded above by 
$$
(1-\tau)\int_{0}^{\upsilon}\sup_{\bx \in \text{supp}(\bX)}\left|G(s|\bx)-G_C(s|\bx)\right|\mathrm ds \leq (1-\tau)\,\upsilon \sup_{\bx \in \text{supp}(\bX)}\sup_{y \leq \upsilon}\left|G(y|\bx)-G_C(y|\bx)\right|,
$$
which converges to 0 when $d_{\mathcal{G}}(G,G_C)\rightarrow 0,$ provided assumption \ref{cond:Ghat} holds. This completes the proof that (A5) holds in our framework. Hence the assumptions of Theorem 1 in DVK are met, from which the weak consistency of $\widehat{\beta}_\tau$ follows. 
\end{proof}

\bibliographystyle{plainnat}
\bibliography{Biblio}	
\end{document}